\documentclass[journal=jctcce,manuscript=article]{achemso}
\setkeys{acs}{articletitle = true}
\usepackage{graphicx}
\usepackage{amsmath}
\usepackage{multirow}
\usepackage{array}
\usepackage{fixltx2e}
\title[]{Linear-scaling local natural orbital CCSD(T)
approach for open-shell systems: algorithm, benchmarks, and large-scale applications}

\author{P. Bern\'at Szab\'o}
\affiliation{Department of Physical Chemistry and Materials Science,
Faculty of Chemical Technology and Biotechnology,
Budapest University of Technology and Economics,
M\H uegyetem rkp. 3., H-1111 Budapest, Hungary}

\author{J\'ozsef Cs\'oka}
\affiliation{Department of Physical Chemistry and Materials Science,
Faculty of Chemical Technology and Biotechnology,
Budapest University of Technology and Economics,
M\H uegyetem rkp. 3., H-1111 Budapest, Hungary}
\alsoaffiliation{HUN-REN-BME Quantum Chemistry Research Group,
M\H uegyetem rkp. 3., H-1111 Budapest, Hungary}
\alsoaffiliation{MTA-BME Lend\"ulet Quantum Chemistry Research Group,
M\H uegyetem rkp. 3., H-1111 Budapest, Hungary}

\author{Mih\'aly K\'allay}
\affiliation{Department of Physical Chemistry and Materials Science,
Faculty of Chemical Technology and Biotechnology,
Budapest University of Technology and Economics,
M\H uegyetem rkp. 3., H-1111 Budapest, Hungary}
\alsoaffiliation{HUN-REN-BME Quantum Chemistry Research Group,
M\H uegyetem rkp. 3., H-1111 Budapest, Hungary}
\alsoaffiliation{MTA-BME Lend\"ulet Quantum Chemistry Research Group,
M\H uegyetem rkp. 3., H-1111 Budapest, Hungary}

\author{P\'eter R. Nagy}\email{nagy.peter@vbk.bme.hu}
\affiliation{Department of Physical Chemistry and Materials Science,
Faculty of Chemical Technology and Biotechnology,
Budapest University of Technology and Economics,
M\H uegyetem rkp. 3., H-1111 Budapest, Hungary}
\alsoaffiliation{HUN-REN-BME Quantum Chemistry Research Group,
M\H uegyetem rkp. 3., H-1111 Budapest, Hungary}
\alsoaffiliation{MTA-BME Lend\"ulet Quantum Chemistry Research Group,
M\H uegyetem rkp. 3., H-1111 Budapest, Hungary}

\date{\today}

\begin{document}

\begin{abstract}

The extension of the highly-optimized local natural
orbital (LNO) coupled-cluster (CC) with single-, double-, and
perturbative triple excitations [LNO-CCSD(T)] method is presented for 
high-spin open-shell molecules based on restricted open-shell references.
The techniques enabling the outstanding efficiency of the closed-shell 
LNO-CCSD(T) variant are adopted, including the iteration- and redundancy-free MP2 and (T) formulations, as well as the
integral-direct, memory- and disk use economic, and OpenMP-parallel algorithms. 
For large molecules, the efficiency of our open-shell LNO-CCSD(T) method
approaches that of its closed-shell parent method due to the application
of restricted orbital sets for demanding integral
transformations and a novel approximation for higher-order long-range spin-polarization effects.
The accuracy of open-shell LNO-CCSD(T) is extensively tested for 
radicals and reactions thereof, ionization processes, 
as well as spin-state splittings and transition-metal compounds.
At the size range, where the canonical CCSD(T) reference is accessible (up to 20--30 atoms)
the average open-shell LNO-CCSD(T) correlation energies are found to be 99.9--99.95\% accurate, which translates into
average absolute deviations of a few tenth of a kcal/mol in the investigated energy differences already with the default settings. 
For more extensive molecules the local errors may grow but they can be estimated and decreased via affordable systematic convergence studies.
This enables the accurate modeling of large systems with complex electronic structure, as illustrated 
on open-shell organic radicals and transition metal complexes of up to 179 atoms, as well as
on challenging biochemical systems, including 
up to 601 atoms and 11,000 basis functions. While the protein models involve difficulties for local approximations, such as
the spin states of a bounded iron ion or an extremely delocalized singly occupied orbital, 
the corresponding single-node LNO-CCSD(T) computations were feasible in a matter of days with 10s to a 100 GB of memory use.
Therefore, the new LNO-CCSD(T) implementation enables highly-accurate computations 
for open-shell systems of unprecedented size and complexity with widely accessible hardware. 

\end{abstract}

\section{Introduction}

The accurate modeling of open-shell species remains challenging 
due to their potentially complicated electronic structure.
Among those, the systems of interest here  exhibit a high-spin open-shell ground state wave function,
which can usually be described by single-reference quantum chemical methods.
Such a wide range of systems include  radicals 
appearing as stable species, or intermediates and transition states of reactions, 
products of ionization or electron attachment processes, etc. often
relevant in redox, atmospheric, polymer, combustion, astro-,  and electrochemistry, just 
to name a few representative fields.\cite{OSissuePCCP}

While density functional theory (DFT) based methods remain the workhorse of computational chemistry due to their relatively affordable cost, 
cases posing challenges for DFT approaches occur more often for open-shell than for closed-shell systems.
Therefore both the necessity and the difficulties of utilizing the wave function based 
treatment of electron correlation were extensively studied also for open-shell species\cite{OSreviewKrylov,OSproblemsStantonGauss,OSQCintro}.
While  the 
second-order M\o ller--Plesset  (MP2)\cite{MP2} approach remains popular due to its cost-efficiency and relative simplicity, 
coupled-cluster (CC) methods represent a faster converging series to the chemically accurate (within 1~kcal/mol) 
description of processes of single-reference species.\cite{RJBRevModPhys,CrawfordRev}
In particular, the ``gold standard'' CC model with single, double (CCSD), and perturbative triple excitations [CCSD(T)],\cite{CCSD(T)}
offers a good compromise between computational-cost and robust accuracy.

Still, the steep $\mathcal{O}({\cal N}^4)$- and $\mathcal{O}({\cal N}^7)$-scaling
data storage and operation count complexity of CCSD(T) with system size ${\cal N}$ limits its applicability range 
to molecules of up to 20--25 atoms. This is an even more severe problem for open-shell species, where unrestricted CC formalisms require
the solution of about three times as many equations 
as their restricted counterparts. 
Moreover, the higher technical complexity of the unrestricted CC methods also slows down the adaptation of new
approaches proposed frequently only for the spin-restricted case. 
For example, density-fitting (DF, or resolution-of-identity) approaches can 
help to deal with  the $\mathcal{O}({\cal N}^4)$-scaling storage complexity\cite{qchem-ri-ccsd,PSI4DF-CCSD(T),mpqcccsdpt,FHI-aims,MPICCSDpT},
while efficient parallelization can reduce the wall time  (but not the scaling) of CCSD(T)\cite{nwchemCCSD(T)1,CCSD(T)molcas,GAMESSUSCCSD(T),aquarius,mpqcccsdpt,FHI-aims,MPICCSDpT}. However, 
even the combination of these techniques with additional advancements in our recent integral-direct, parallel DF-CCSD(T) implementation 
pushes the limits of conventional CCSD(T) computations only up to 30 atoms.\cite{MPICCSDpT} 

This motivates the introduction of reduced-scaling 
approximations, such as the robust frozen natural orbital (NO) approach\cite{NOLowdin,NOBartlett}, which can 
extend the applicability range of NO-CCSD(T) somewhat further.\cite{FNONAF,FNOExt} 
Alternatively, one can also accelerate the basis set convergence via 
explicitly correlated (F12)  CC methods \cite{R12,R12Rev,F12ChemRev} leading to more compact
atomic orbital (AO) basis set requirements.
However, even our recent combination of the NO approach and an optimized F12 implementation\cite{F12pT} 
 allowed us to approach the complete  basis set (CBS) limit for closed-shell species of up to 50 atoms with an  
even smaller open-shell limit of about 30--35 atoms.\cite{RedCostF12}

At this size range, it starts to be beneficial to take advantage 
of the rapid decay of electron correlation with the distance via local correlation
 approaches.\cite{linearscalebook,fragmentreviewHerber19,fragmentbook}
Current methods still build on  fundamental techniques pioneered
by Pulay and Saeb{\o }\cite{PulayLocalCEPA,PulayLocalMP2}, such as
the approximation of the energy contribution of distant localized molecular orbital (LMO) pairs (pair approximation),
and the restriction of the correlating orbital spaces to a spatially compact lists surrounding 
the strongly interacting LMOs  (domain approximation).
One group of methods then approximates or neglects the distant pair interactions 
leading to a number of decoupled subsystem MP2 or CC equations to be solved,
but here, it remains challenging to deal with the significant overlap of the subsystems.\cite{linearscalebook,fragmentreviewHerber19,EBfragmentReview,fragmentreview15} 
Advanced representatives of this strategy also developed up to the CC level include 
the cluster-in-molecule (CIM) approach of Li, Li, Piecuch, Guo, and their co-workers,\cite{PiecuchLocalCC,CIM-review,CIM-DLPNO}
the divide-expand-consolidate (DEC) approach of J{\o}rgensen, Kristensen, and their co-workers,\cite{DEC,DEC-CCSD(T)}
the divide-and-conquer (DC) method of Li and Li\cite{DC1} and Kobayashi and Nakai,\cite{DC-CC,DC-HOCC}
the incremental method proposed by Stoll\cite{Incremental1} and 
further employed in the local correlation context by Friedrich and Dolg.\cite{IncrementalSymm,IncrCCSD(T),IncrPNO}

One can also take advantage of the wave function sparsity
not only in real space but also via NO-based data
compression approaches.\cite{OSV1,DLPNO_CCSD_OS,OS_DLPNO-CCSD(T),PNOCCreview,PNO-UCCSD(T),LT-CCSD(T)-F12,LocalCC3,LocalCC4}
Among these, the LMO pair specific pair natural orbitals (PNOs) are 
employed in the domain based local PNO (DLPNO) method of Neese, Valeev and co-workers,\cite{DLPNO-CCSDiterativeT,NeeseOSPNO,DLPNO_CCSD_OS,OS_DLPNO-CCSD(T),DLPNO_CCSD(T)-F12_OS}
but PNO-based approaches were also developed up to the CCSD(T) level 
by Werner and Ma\cite{PNOCCreview,PNO-UCCSD,PNO-UCCSD(T)}, as well as
by H\"attig and Tew.\cite{LT-CCSD(T)-F12}
Compared to those, our local natural orbital (LNO) family of methods employs an LMO-specific NO set 
compressing both the occupied and virtual orbital spaces.\cite{LocalCC,LocalCC2,LocaldRPA,RedScalADC,LaplaceT,LocalCC3,MPICCSDpT,LocalCC4}

Due to the considerable challenges associated with open-shell systems and unrestricted CC formalisms,
fewer local CCSD(T) methods are available for systems other than those of a closed-shell singlet electronic structure. 
The incremental scheme was extended to both unrestricted Hartree--Fock (UHF)\cite{FriedrichUHF} as well as 
restricted open-shell HF (ROHF)\cite{DualIncrOS} references.
Additionally, the high-spin open-shell variants of the PNO-L
methods by Werner and  Ma\cite{PNO-UCCSD,PNO-UCCSD(T)}, as well as the DLPNO method
by Neese, Valeev, Hansen, Saitow, Guo, Kumar, and co-workers\cite{NeeseOSPNO,DLPNO_CCSD_OS,OS_DLPNO-CCSD(T),DLPNO_CCSD(T)-F12_OS}
were also introduced recently, while here, we present the high-spin open-shell extension of our LNO-CCSD(T) approach.

To that end, here, we  combine the attractive properties of two lines of recent  developments 
within our restricted  LNO-CCSD(T)\cite{LaplaceT,LocalCC3,LocalCC4} as well as our open-shell local MP2 (LMP2)\cite{ROLocalMP2} schemes.
The outstanding efficiency of these approaches originates from the 
Laplace-transform based, redundancy-free evaluation of the amplitudes both at the LMP2 and the (T) level.\cite{ROLocalMP2,LaplaceT}
Moreover, the CCSD contribution is also obtained in the compact LNO space, which was further accelerated 
via our highly-optimized CCSD implementation designed also for the unconventional ratio of the occupied and virtual orbital dimensions occurring in the LNO basis.\cite{MPICCSDpT}
The resulting LMP2 and LNO-CCSD(T) algorithms are fully ab initio, i.e., free from empirical or distance based
cutoff parameters, manual fragment definitions, bond cutting and capping, etc. usually associated with local correlation approaches.
In fact, the LNO approximations are defined completely automatically and
adopt to the complexity of the wave function of the systems. Exploiting this property, 
we designed an LNO approximation hierarchy of threshold combinations (e.g., Normal, Tight, $\dots$),
which form a systematically convergent series usually also suitable for 
extrapolation toward conventional CCSD(T) and for providing a conservative LNO error estimate.\cite{LocalCC4}
Further unique features of our LMP2 and LNO-CCSD(T) methods
include the exceptionally small memory and disk use,
checkpointing, utilization of  point group symmetry (even non-Abelian) 
and treatment of near-linear dependent AO basis sets.\cite{LocalMP2,LocalCC3,LocalCC4}
These capabilities were all required in our largest local CCSD(T) computation so 
far performed for extended supramolecular complexes\cite{L7CCDMC}
as well as for a protein of 1023 atoms with almost 45,000 AOs in a quadruple-$\zeta$ basis set.\cite{LocalCC4}
Even larger systems can be targeted via various embedding approaches combining LNO-CC based chemically active 
regions embedded into lower-level LNO-CC, LMP2, DFT and/or molecular mechanics (MM) environments.\cite{Embed,Dual}


An important goal in the generalization of the above spin-restricted local correlation methods 
to open-shell systems is to retain as much as possible the computational efficiency of the closed-shell scheme.
Therefore, in our open-shell LMP2 method\cite{ROLocalMP2} as well as here 
for the LNO-CCSD(T) case, we employ a restricted open-shell (RO) reference determinant, RO LMO set,
and RO intermediate basis sets used for the costly integral transformation steps. 
Interestingly, this strategy is implemented in three completely different manners
in the two PNO-based and our approach. The DLPNO method 
 employs
an $n$-electron valence state perturbation theory Ansatz,\cite{DLPNO_CCSD_OS}
the PNO-L methods utilize a spin-adapted MP2 formulation,\cite{PNO-UCCSD}
 while we  employ a spin-restricted integral transformation combined with a simpler 
ROHF-based but unrestricted MP2 Ansatz\cite{ROHFMP2,ROHFMP2_Knowles}. 
Building on that, here we construct a restricted LNO basis. However, due to the properties 
of the perturbative triples corrections, in the end, both the PNO and LNO methods 
have to use unrestricted formulae for the CCSD(T) part. 

To enable the cost of the unrestricted  CCSD(T) calculations of the LNO scheme to
approach that of the closed-shell method,  at least in the asymptotic limit, we utilize 
an additional approximation that we developed for our open-shell LMP2 approach.\cite{ROLocalMP2}
 Briefly, we can exploit that the long-range spin-polarization 
effects of localized singly occupied MOs (SOMOs) can be taken into account at the mean-field and approximated MP2 level.\cite{ROLocalMP2}
 Then, the efficient closed-shell formulae and algorithms 
can be utilized for the LMP2 and LNO-CCSD(T)  correlation energy contributions of the LMOs that 
are not interacting strongly with any SOMOs. Interestingly,  here, we also find that for large systems of about 200 atoms or more, 
up to 50--90\% of the LMO correlation energy contributions can be safely  evaluated with this approach 
at practically the cost of the closed-shell counterpart. 

The capabilities of the resulting open-shell LNO-CCSD(T) code are illustrated on three-dimensional 
transition metal complexes of up to 179 atoms, as well as on
protein models of 565 and 601 atoms. 
The protein models  involve about twice as many (cca.~11,000) AOs as
the largest open-shell local CCSD(T) computations in the literature so far.\cite{OS_DLPNO-CCSD(T),DLPNO_CCSD(T)-F12_OS}
In addition to the more complicated electronic structure of the studied metal-complexes, 
the largest protein model also exhibits a highly delocalized SOMO posing a challenge to any local correlation approach. 
Nevertheless, these LNO-CCSD(T) computations were still feasible within  wall times of about 2--4 days
using a single CPU with 20 physical cores 
and mostly a few tens to at most 100 GBs of memory and comparable disk use. 

It is even more important to retain the accuracy of restricted LNO-CCSD(T) for the 
more challenging open-shell applications. The closed-shell LNO-CCSD(T) method 
was extensively benchmarked by us\cite{LocalCC3,LocalCC4,MichaelJACS,MeLysine,L7CCDMC,S66CC,phosphinyl} 
as well as independently\cite{MobiusCC,LCCcompRuMartin,Paulechka,LCC_MOBH35re,LCCcomp_anionbind,S66x8LCCMartin,ACONF_LCCMatrin,fullererneLCCiso} 
 revealing highly competitive accuracy and efficiency compared to other local CCSD(T) approaches, e.g., 
for  organic thermochemistry,\cite{LocalCC4,Paulechka}
non-covalent complexes,\cite{L7CCDMC,S66CC,S66x8LCCMartin} 
conformational and isomerization energies,\cite{ACONF_LCCMatrin,fullererneLCCiso}
ionic interactions\cite{LCCcomp_anionbind,MeLysine,IonLigNa,IonLigMg}
as well as for organometallic-\cite{LCCcompRuMartin,LCC_MOBH35re}
 and  extended $\pi$-systems\cite{MobiusCC,L7CCDMC}
exhibiting even moderate non-dynamic correlation.
Here, we extend these benchmarks to radical stabilization energies, ionization potentials, and spin-state energies of small- to medium-sized systems
and up to triple-$\zeta$ basis sets. On the average, at the range where we can compare against the canonical CCSD(T) reference (that is, ca. 20--30 atoms),
 we find the open-shell LNO-CCSD(T) correlation energies 
99.9--99.95\% accurate, which translates into a few tenth of a kcal/mol average energy difference deviations for the investigated systems. 
This is in accord with the accuracy of the closed-shell method, 
where, however, one should point out that the local errors somewhat grow with increasing system size 
and wave function complexity.\cite{MichaelJACS,L7CCDMC,LocalCC3,LocalCC4} 
In practice, the LNO error can be estimated and  systematically converged close to the 
local approximation free limit at an affordable cost as previously demonstrated in various applications.\cite{LocalCC3,LocalCC4,MichaelJACS,MeLysine,L7CCDMC,S66CC,phosphinyl}

The discussion of the corresponding details is organized as follows. 
Sections \ref{sec:theory} and \ref{sec:algorithm} collect the theoretical and algorithmic 
details of the new LNO-CCSD(T) approach focusing on the 
technicalities emerging specifically for the open-shell case. 
The computational details and the benchmark molecules 
are introduced in Section \ref{sec:compdet}. 
The accuracy of the individual and combined local approximations is assessed
in Sections \ref{sec:convergence} and \ref{sec:statistics}.
Finally, large-scale applications for systems of 175--601 atoms and the
corresponding computational
requirements are discussed in Section \ref{sec:large_calcs}.


\section{Theoretical background}\label{sec:theory}

Throughout the presented derivations, restricted open-shell (RO) reference determinants are assumed, consisting of singly and doubly occupied molecular orbitals (SOMOs and DOMOs, respectively).
Since the LNO method makes use of multiple orbital types, 
the notation of these is summarized in Table \ref{tab:orb_notation}.
The conventional and the LNO 
correlation energy expressions also employ 
unrestricted, semi-canonical (also known as pseudo-canonical)
MOs. The lower (upper) case indices label orbitals with spin up (down) 
occupation, while $i,j,k\dots, I,J,K,\ldots$ and $a,b,c,\ldots, A,B,C,\ldots$
indices are used for the occupied and virtual subsets, respectively.  
Local approximations are introduced in the basis of
localized molecular orbitals (LMOs) obtained 
from a restricted open-shell reference, which will be denoted in general by indices $\mathcal{I},\mathcal{J},\mathcal{K}, \ldots$, 
while these LMOs  will be labeled as $i^{\prime},j^{\prime},k^{\prime},\ldots$ 
($I^{\prime},J^{\prime},K^{\prime},\ldots$), respectively, when  occupied   by spin up (spin down) electrons.
\begin{table}
\caption{Summary of index notations for orbital sets employed in Sections \ref{sec:theory} and \ref{sec:algorithm}.}
\small
\begin{tabular}{cl}\hline
$i,j,k,\ldots$ ($I,J,K,\ldots$) & spin up (spin down) (semi-)canonical occupied orbitals\\
$a,b,c,\ldots$ ($A,B,C,\ldots$) & spin up (spin down) (semi-)canonical virtual orbitals  \\
$i^{\prime},j^{\prime},k^{\prime},\ldots$ ($I^{\prime},J^{\prime},K^{\prime},\ldots$) & spin up (spin down) localized restricted occupied orbitals\\
$\mathcal{I},\mathcal{J},\mathcal{K}, \ldots$ & localized restricted occupied orbitals (spatial) \\
$\hat i,\ldots, \hat a,\ldots$ & restricted orbitals in the extended domain \\
$\tilde i,\ldots, \tilde a,\ldots$ ($\tilde I,\ldots, \tilde A,\ldots$) & spin up (spin down) (semi-)canonical orbitals in primary/extended domains  \\
$\bar i,\ldots, \bar a,\ldots$ & restricted orbitals in the local interacting subspace \\
$\underline i,\ldots, \underline a,\ldots$ ($\underline I,\ldots, \underline A,\ldots$) & spin up (spin down) (semi-)canonical orbitals in the local interacting subspace \\
$\mu, \nu, \lambda \ldots$ & atomic orbitals \\
$X, Y, \ldots$ & auxiliary functions for the DF approximation 
\end{tabular}
\label{tab:orb_notation}
\end{table}

\subsection{Open-Shell LNO-CCSD(T) Ansatz}

Following the relevant approaches introduced for the previous members of the LNO family 
of methods,\cite{LocalCC,LocalCC2,LocalMP2,LaplaceT,LocalCC3,LocalCC4,RedScalADC,ROLocalMP2}
 especially the open-shell LMP2\cite{ROLocalMP2} and the closed-shell LNO-CCSD(T),\cite{LaplaceT,LocalCC3,LocalCC4,MPICCSDpT}
here, we introduce the open-shell LNO-CCSD(T) Ansatz built on restricted open-shell references. 
First, conventional unrestricted CCSD and (T) energy expressions\cite{DPD,CCSD(T)} 
 are written in terms of semicanonical orbitals suitable for 
transformations to the (restricted open-shell) LMO basis due to their invariance to unitary orbital rotations.

To introduce the orbital-specific correlation energy 
contributions of the LNO approach, first the open-shell CCSD correlation energy expression is rewritten as a sum of contributions from the occupied spin up and spin down orbitals, $\delta E^{\mathrm{CCSD}}_{i}$ and $\delta E^{\mathrm{CCSD}}_{I}$:
\begin{equation}
\begin{split}
    E^{\mathrm{CCSD}} =& \sum_{i}\delta E^{\mathrm{CCSD}}_{i} + \sum_{I}\delta E^{\mathrm{CCSD}}_{I} \\
                      =&\sum_{i}\left(\sum_{a} t^{a}_{i}f^{a}_{i} + \frac{1}{4} \sum_{abj}\tau^{ab}_{ij}\langle ab \Vert ij \rangle+\frac{1}{2}\sum_{aBJ}\tilde \tau^{aB}_{iJ}\langle aB | iJ \rangle\right) \\
&+\sum_{I}\left(\sum_{A} t^{A}_{I}f^{A}_{I} + \frac{1}{4} \sum_{ABJ}\tau^{AB}_{IJ}\langle AB \Vert IJ \rangle+\frac{1}{2}\sum_{Abj}\tilde \tau^{Ab}_{Ij}\langle Ab | Ij \rangle\right) \, ,
\end{split}
\label{eqn:cim_ccsd_energy}
\end{equation}
where $\tau^{ab}_{ij} = t^{ab}_{ij} + t^{a}_{i}t^{b}_{j} - t^{b}_{i}t^{a}_{j}$, $\tilde \tau ^{aB}_{iJ} = t^{aB}_{iJ} + t^{a}_{i}t^{B}_{J}$, and $t$  denotes the CCSD singles and doubles cluster amplitudes.
 Additionally, $\langle ab \Vert ij \rangle$ stands for antisymmetrized electron repulsion integrals (ERIs), constructed as $\langle ab \Vert ij \rangle = \langle ab \vert ij \rangle - \langle ab \vert ji \rangle$ using the Dirac notation. 

Analogously, the energy formula for the open-shell perturbative triples correction of CCSD(T) 
can be written as
\begin{equation}
\begin{split}
    E^{\mathrm{(T)}} =& \sum_{i}\delta E^{\mathrm{(T)}}_{i} + \sum_{I}\delta E^{(T)}_{I} \\
=&\sum_{i}\left(\frac{1}{36}\sum_{abcjk}t^{abc}_{ijk}W^{abc}_{ijk}+\frac{1}{8}\sum_{aBCJK}t^{aBC}_{iJK}W^{aBC}_{iJK}+\frac{1}{8}\sum_{abCjK}t^{abC}_{ijK}W^{abC}_{ijK}\right) \\
&+\sum_{I}\left(\frac{1}{36}\sum_{ABCJK}t^{ABC}_{IJK}W^{ABC}_{IJK}+\frac{1}{8}\sum_{Abcjk}t^{Abc}_{Ijk}W^{Abc}_{Ijk}+\frac{1}{8}\sum_{ABcJk}t^{ABc}_{IJk}W^{ABc}_{IJk}\right) \, ,
\end{split}
\label{eqn:cim_pt_energy}
\end{equation}
where $t^{abc}_{ijk}$ denotes the triple excitation amplitude of the (T) method, and $W^{abc}_{ijk}$ can be 
written as $t^{abc}_{ijk} D^{abc}_{ijk} $, that is, the triple excitation amplitude multiplied by 
the canonical orbital energy differences collected into the denominator $D^{abc}_{ijk} $.\cite{CCSD(T),CrawfordRev}
Combining eq \ref{eqn:cim_ccsd_energy} and eq \ref{eqn:cim_pt_energy}, the full CCSD(T) correlation energy can also be written as a sum of orbital contributions:
\begin{equation}
\begin{split}
E^{\mathrm{CCSD(T)}} =& \sum_{i} \delta E^{\mathrm{CCSD(T)}}_{i} + \sum_{I} \delta E^{\mathrm{CCSD(T)}}_{I} \\
=& \sum_{i} \left(\delta E^{\mathrm{CCSD}}_{i} + \delta E^{\mathrm{(T)}}_{i}\right) +  \sum_{I} \left(\delta E^{\mathrm{CCSD}}_{I} + \delta E^{\mathrm{(T)}}_{I}\right) \, .
\label{eqn:cim_ccsdpt_energy}
\end{split}
\end{equation}

To introduce the 
 local CCSD(T) Ansatz, let us transform the separated occupied indices of eq \ref{eqn:cim_ccsdpt_energy} to 
the restricted open-shell LMO basis:
\begin{equation}
E^{\mathrm{CCSD(T)}} = \sum_{i^\prime}\delta E^{\mathrm{CCSD(T)}}_{i^\prime} + \sum_{I^\prime}\delta E^{\mathrm{CCSD(T)}}_{I^\prime} = \sum_{\mathcal I}\delta E^{\mathrm{CCSD(T)}}_{\mathcal I} \, ,
\label{eqn:loc_ccsdpt_energy}
\end{equation}
where in the last term, $\mathcal{I}$ denotes a spatial orbital occupied by either one or two electrons 
in the RO LMO basis,
while $i^{\prime} $ ($I^{\prime} $) refers to
 orbitals with the same spatial component as LMO $\mathcal{I}$, 
but occupied by at most one spin up (spin down) electron. The explicit expressions used for the LNO
method will be introduced in Sect.~\ref{sect:lisE}. 

The scaling of CCSD(T)  can be made asymptotically linear with respect to the system size
if only a domain of asymptotically constant number of orbitals is required 
to evaluate the correlation energy contribution of a given orbital $\mathcal I$, 
that is, $\delta E^{\mathrm{CCSD(T)}}_{\mathcal I}$. 
To construct the domain around each LMO, which is then called the central LMO of its own domain, 
first, those LMOs are collected that strongly interact with the central LMO.
The selection of strongly interacting LMO pairs 
 is based on multipole approximated MP2 pair energies [see the $\delta E^{\mathrm{MP2}}_{\mathcal{IJ}}\left(\mathcal{P}_\mathcal{IJ}\right)$ term of eq \ref{eqn:mp2_correction}]
 evaluated in LMO pair specific domains (PD, $\mathcal{P}_\mathcal{IJ}$). 
LMO pairs with very small $\delta E^{\mathrm{MP2}}_{\mathcal{IJ}}\left(\mathcal{P}_\mathcal{IJ}\right)$
pair correlation energy estimates are considered distant and, unlike to the strong pairs,
 they do not enter to the higher-level computations. 
Next, the virtual space of the domain of LMO $\mathcal I$ is constructed 
from local  projected atomic orbitals (PAOs) surrounding the 
central LMO and its strongly interacting LMO pairs.  
The domain obtained in this way is referred to as the extended domain (ED, $\mathcal{E}_{\mathcal I}$) of LMO $\mathcal I$, 
which is sufficiently compact to efficiently perform MP2 computations exploiting our
open-shell LMP2 implementation.\cite{ROLocalMP2}
The resulting local MP2 correlation energy contribution of the ED 
 [$\delta E^{\mathrm{MP2}}_{\mathcal I}\left(\mathcal{E}_{\mathcal I}\right)$] is utilized as part of the 
 correction employed to decrease the effect of the remaining approximations in the ED at the CCSD(T) level
(see eq \ref{eqn:mp2_correction}).
To further compress the orbital spaces of the ED, LNOs
are constructed using the density built from the first-order M\o ller--Plesset (MP1) amplitudes of the ED, yielding 
the local interacting subspace (LIS, $\mathcal{L}_{\mathcal I}$) of LMO $\mathcal I$.
Consequently, the LNO-CCSD(T) correlation energy reads as
\begin{equation} 
E^{\mathrm{LNO-CCSD(T)}} = \sum_{\mathcal I}\left[ \delta E^{\mathrm{CCSD(T)}}_{\mathcal I} \left(\mathcal L_{\mathcal I}\right) + \Delta E^{\mathrm{MP2}}_{\mathcal I}\right] \, ,
\label{eqn:lno_ccsdpt_energy}
\end{equation}
where the energy correction $\Delta E^{\mathrm{MP2}}_{\mathcal I}$ is calculated at the MP2 level of theory as
\begin{equation}
\Delta E^{\mathrm{MP2}}_{\mathcal I} = \delta E^{\mathrm{MP2}}_{\mathcal I}\left(\mathcal E_{\mathcal I}\right) - \delta E^{\mathrm{MP2}}_{\mathcal I} \left(\mathcal L_{\mathcal I}\right) + \frac{1}{2} \sum^{\mathrm{distant}}_{\mathcal J} \delta E^{\mathrm{MP2}}_{\mathcal{IJ}}\left(\mathcal P_{\mathcal{IJ}}\right) \, .
\label{eqn:mp2_correction}
\end{equation}
Therefore, correlation energy terms are included in the final LNO-CCSD(T) expression for all orbital pairs of the entire molecule. 
The largest and most important component, which corresponds to the correlation of the strong pairs,
is included at the complete CCSD(T) level in the LISs. The contribution of the frozen LNOs
and the correlation energy contribution of distant LMO pairs 
are included at the MP2 level in the extended and pair domains, respectively. 




\section{Algorithm}\label{sec:algorithm}
\begin{figure}
\centering
\includegraphics{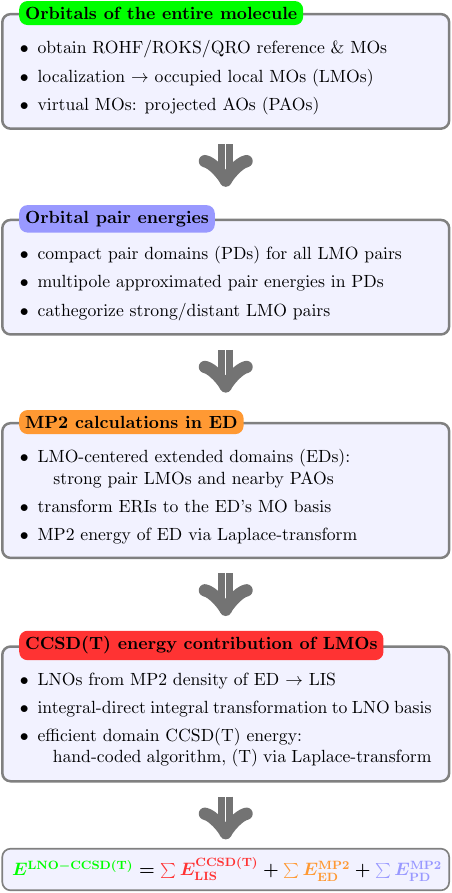}
\caption{Major algorithmic steps of the presented restricted open-shell LNO-CCSD(T) method.}
\label{fig:algorithm}
\end{figure}

The algorithm of the restricted open-shell LNO-CCSD(T) method is summarized in Figure \ref{fig:algorithm} and described in this section step by step in detail.

\subsection{Self-Consistent Field and Orbital Localization}\label{sec:scf}

The considerations regarding the reference selection
 are analogous to those of our local RO MP2 scheme.\cite{ROLocalMP2}
The restricted open-shell  Hartree-Fock (HF) or Kohn-Sham (KS) reference determinant
can be obtained via a restricted open-shell self-consistent field (SCF) calculation (ROHF/ROKS), 
or via unrestricted  (UHF/UKS) computations  followed by constructing quasi-restricted orbitals (QROs).\cite{QRO,ROLocalMP2}

The computational cost of (RO/U)HF  can become  high above
 the few hundred atom range even with density-fitting (DF) approaches. 
To reduce the scaling of conventional DF-SCF calculations, for example, 
the local DF approach can be utilized, which restricts the lists of auxiliary functions and, for very large systems, also
that of AOs at the exchange matrix evaluation 
to include functions that are spatially close to an LMO \cite{LocDF-HF,LocalMP2,Multipole,DualDF}. 
In the present study, local DF is employed only for the ROHF and UHF computations in the 500- to 600-atom range.

The occupied orbitals of the reference determinant are localized with the algorithm proposed by Boys \cite{BoyLoc2},
but Pipek--Mezey\cite{PipekMezey}, intrinsic bond orbital\cite{IBO}, and generalized Boys methods 
with higher orbital variance power\cite{JansikJCP} are also implemented. 
The localization is carried out in a spin restricted manner, and the 
doubly occupied and open-shell MOs are not mixed resulting in doubly and singly occupied LMOs.
The clear advantage of the restricted formalism 
is that the correlation energy contribution of the spin up and down electrons of a 
doubly occupied LMO can be obtained in a single domain corresponding to a RO LMO.
The alternative way of computing 
$\delta E^{\mathrm{CCSD(T)}}_{i^\prime}$ and $ \delta E^{\mathrm{CCSD(T)}}_{I^\prime}$ 
of eq \ref{eqn:loc_ccsdpt_energy} in separate, spin case dependent domains would be 
approximately twice as costly. 
The potential drawback of keeping the LMO space restricted 
is that in systems with only one or a few SOMOs (or when the mixing of the SOMOs is prohibited, e.g., due to 
symmetry), it may not be possible to (sufficiently) localize the SOMO(s), leading to singly occupied orbitals
in the LMO basis that are  (potentially) still delocalized (see below an example in Fig. \ref{fig:daao_large_orbital}). 


\subsection{PAO construction and pair energy calculation}

The PAO construction and pair energy calculation algorithms
follow closely those of the open-shell LMP2 scheme\cite{ROLocalMP2} with minor extensions needed for the LNO-CC methods discussed here in more detail.
To identify the strongly interacting LMO pairs, their approximate MP2 pair correlation energies are computed in LMO pair specific domains (PDs).
The PDs contain the corresponding occupied LMO pair and PAOs as virtual functions
that are centered on atoms surrounding the two LMOs.

To assemble the virtual space, first, the PAOs of the entire molecule are constructed
by projecting out the restricted LMOs (both DOMOs and SOMOs) from all AOs of the molecule ($\vert \mu \rangle$):
\begin{equation}
\vert a_{\mu} \rangle = \left( 1 - \sum^{\mathrm{DO} \, \cup \, \mathrm{SO}}_{\mathcal I} \vert \mathcal I \rangle \langle \mathcal I \vert \right) \vert \mu \rangle \, .
\label{eqn:pao}
\end{equation}
The AO $\vert \mu \rangle$ and the corresponding atom can be considered as the center of the resulting PAO $\vert a_{\mu} \rangle$.
The PAOs of eq \ref{eqn:pao} are restricted and span the virtual subspace of the spin up electrons.
To span the virtual subspace of the spin down electrons, we use the union of all SOMOs and PAOs.

The pair domain of a given LMO pair is constructed as the union of the primary domains of the two LMOs.
To define the  primary domains, a set of atoms is assigned to each LMO and PAO using a modified \cite{LocalMP2,LocalCC3} Boughton--Pulay (BP) algorithm \cite{Boughton}.
The BP atom list of an LMO or PAO is compiled such that the projection
 of the LMO/PAO onto the AOs of its BP atom list
 has an overlap value of at least $T$ with the original, unprojected LMO/PAO.
Thus, $1-T$ gives an upper bound for the truncation error of the projection onto the BP atom lists. 
For the  primary domain  construction, BP atom lists are assembled
 using $T_{\mathrm{PDo}}=0.999$ for LMOs and $T_{\mathrm{PDv}}=0.98$ for PAOs.
The occupied subspace of the  primary domain  contains a single LMO, 
while its virtual subspace includes those PAOs that are centered on any of the atoms in the LMO's BP atom list.
Additionally, if a BP list of a SOMO constructed with completeness criterion $T_{\mathrm{PDv}}$ 
overlaps with the BP set of the LMO, the SOMO is also included in the spin down virtual subspace of the primary domain.
Then, the atom list of the PD contains the union of  the BP atom lists of all orbitals in the primary domain.
Finally, the orbitals of the primary domain  are projected onto the AOs of the  primary domain,
 the projected orbitals are orthogonalized, and the spin up and spin down orbitals are separately canonicalized.

The multipole approximated MP2 pair energy of an LMO pair is then evaluated in the primary domains' bases 
of the LMO pair:
\begin{equation}
\begin{split}
\delta E^{\mathrm{MP2}}_{\scriptscriptstyle \mathcal{IJ}}\left(\mathcal{P_{\scriptscriptstyle \mathcal{IJ} }}\right) =
\delta E^{\mathrm{MP2}}_{i^{\prime} j^{\prime}  }\left(\mathcal{P}_{\scriptscriptstyle \mathcal{IJ}}\right) +
\delta E^{\mathrm{MP2}}_{I^{\prime} J^{\prime}  }\left(\mathcal{P}_{\scriptscriptstyle \mathcal{IJ}}\right) +
\delta E^{\mathrm{MP2}}_{i^{\prime} J^{\prime}  }\left(\mathcal{P}_{\scriptscriptstyle \mathcal{IJ}}\right) +
\delta E^{\mathrm{MP2}}_{I^{\prime} j^{\prime}  }\left(\mathcal{P}_{\scriptscriptstyle \mathcal{IJ}}\right) = \\
-\sum_{\tilde a_{\scriptscriptstyle \mathcal{I}} \, \tilde b_{\scriptscriptstyle \mathcal{J}} }
\frac{[( \tilde a_{\scriptscriptstyle \mathcal{I}} i^{\prime}  \vert  \tilde b_{\scriptscriptstyle \mathcal{J}} j^{\prime}  )^{\left[4\right]} ]^{2}}
{\varepsilon_{\tilde a_{\scriptscriptstyle \mathcal{I} }} + \varepsilon_{\tilde b_{\scriptscriptstyle \mathcal{J} }} - F_{i^{\prime} i^{\prime} } - F_{j^{\prime} j^{\prime} }}
-\sum_{\tilde A_{\scriptscriptstyle \mathcal{I}} \, \tilde B_{\scriptscriptstyle \mathcal{J}} }
\frac{[( \tilde A_{\scriptscriptstyle \mathcal{I}} I^{\prime}  \vert  \tilde B_{\scriptscriptstyle \mathcal{J}} J^{\prime}  )^{\left[4\right]} ]^{2}}
{\varepsilon_{\tilde A_{\scriptscriptstyle \mathcal{I} }} + \varepsilon_{\tilde B_{\scriptscriptstyle \mathcal{J} }} - F_{I^{\prime} I^{\prime}} - F_{J^{\prime} J^{\prime} }} \\
-\sum_{\tilde a_{\scriptscriptstyle \mathcal{I}} \, \tilde B_{\scriptscriptstyle \mathcal{J} }}
\frac{[( \tilde a_{\scriptscriptstyle \mathcal{I}} i^{\prime}  \vert  \tilde B_{\scriptscriptstyle \mathcal{J}} J^{\prime} )^{\left[4\right]} ]^{2}}
{\varepsilon_{\tilde a_{\scriptscriptstyle \mathcal{I} }} + \varepsilon_{\tilde B_{\scriptscriptstyle \mathcal{J} }} - F_{i^{\prime} i^{\prime}} - F_{J^{\prime} J^{\prime} }}
-\sum_{\tilde A_{\scriptscriptstyle \mathcal{I}} \, \tilde b_{\scriptscriptstyle \mathcal{J} }}
\frac{[( \tilde A_{\scriptscriptstyle \mathcal{I}} I^{\prime} \vert  \tilde b_{\scriptscriptstyle \mathcal{J}} j^{\prime} )^{\left[4\right]} ]^{2}}
{\varepsilon_{\tilde A_{\scriptscriptstyle \mathcal{I} }} + \varepsilon_{\tilde b_{\scriptscriptstyle \mathcal{J} }} - F_{I^{\prime} I^{\prime}} - F_{j^{\prime} j^{\prime} }} \, .
\end{split}
\label{eqn:pair_energy}
\end{equation}
Here, the pseudocanonical orbital energy of the virtual orbitals is denoted by $\varepsilon$, while $F_{i^\prime i^\prime}$ ($F_{I^\prime I^\prime}$) indicates a diagonal element of the spin up (spin down) Fock matrix.
The LMO subscript of the virtual orbitals indices indicates that the corresponding summations
 include virtual orbitals only from the primary domain of the given LMO.
The ERIs, denoted here in Mulliken notation by $( \tilde a_{\mathcal{I}}i^\prime \vert \tilde b_{\mathcal{J}} j^\prime )^{[4]}$ are calculated using a multipole expansion up to fourth order, including dipole--dipole, dipole--quadrupole, quadrupole--quadrupole, and dipole--octupole moment terms as discussed in Ref. \citenum{LocalMP2}.

Utilizing the $\delta E^{\mathrm{MP2}}_{\mathcal{IJ}}\left(\mathcal{P}_{\mathcal{IJ}}\right)$  pair 
energies, the $\mathcal{IJ}$ LMO pair can be classified as a strongly interacting pair if $\delta E^{\mathrm{MP2}}_{\mathcal{IJ}}\left(\mathcal{P}_{\mathcal{IJ}}\right) > f_{\mathrm{w}}\varepsilon_{\mathrm{w}}$, 
where $\varepsilon_{\mathrm{w}}$ is the strong pair energy threshold, and $f_{\mathrm{w}}$ is a factor of 1, $\frac{1}{2}$ or $\frac{1}{4}$ for DO--DO, DO--SO and SO--SO LMO pairs, respectively.
The $f_{\mathrm{w}}$ scaling factor is introduced so that pairs between LMOs of different occupations 
 are handled on an equal footing since  SO LMOs have half as many pair
 energy terms in eq \ref{eqn:pair_energy} as DO LMOs.
A more detailed discussion and numerical benchmarks regarding the benefits of using
the  $f_{\mathrm{w}}$  factor are provided in our RO LMP2 study \cite{ROLocalMP2}.
Note that the approximated pair energy of those pairs which are not classified as strong is added to the MP2 correlation energy corrections as the last term of eq \ref{eqn:mp2_correction}.

The pair energies of eq \ref{eqn:pair_energy} have been extensively tested on a large number of (mainly organic) molecules \cite{LocalMP2,LocalCC3,LocalCC4,ROLocalMP2}, with consistently satisfactory performance.
However, the open-shell LNO-CCSD(T) approach is expected to be applied more frequently 
also for transition metal complexes (see, e.g., Sections \ref{sec:convergence} and  \ref{sec:large_calcs}), 
exhibiting electronic structures potentially more complex than that of a typical organic molecule.
In a few complicated cases, we have found that the multipole approximated MP2 pair energies 
can underestimate the approximation-free MP2 pair energies 
 leading to 
 LMO pairs classified as distant instead of strong on the border of the two 
 categories.
To remedy this issue, an additional mechanism is introduced here to extend the strong pair list.
In the present approach, we investigate more closely the orbital pairs characterized originally as distant
 that are on the border of the distant and strong categories.
More precisely, the orbital pairs with
 $\frac{f_{\mathrm{w}}\varepsilon_{\mathrm{w}} }{g_{\mathrm{w} }} < \delta E^{\mathrm{MP2}}_{\mathcal{IJ}}\left(\mathcal{P}_{\mathcal{IJ}}\right) < f_{\mathrm{w}}\varepsilon_{w}$ 
are considered with  parameter $g_\mathrm{w}$ defining the range of pair energies where potentially important 
LMO pairs may still appear. For these pairs, an additional measure is computed:
\begin{equation}
M_{\mathcal{IJ}} = \sqrt{\sum^{\mathrm{atoms}}_{A} \left(
\sum^{\mathrm{on} \, A}_{\mu \nu}C^{A}_{\mu \mathcal I} S^{A}_{\mu \nu} C^{A}_{\nu \mathcal I}
\cdot
\sum^{ \mathrm{on} \, A}_{\mu \nu}C^{A}_{\mu \mathcal J} S^{A}_{\mu \nu} C^{A}_{\nu \mathcal J}
\right)^{2}} \, ,
\label{eqn:mulliken_overlap}
\end{equation}
where, $C^{A}_{\mu \mathcal{I}}$ is an orbital coefficient of truncated LMO $\mathcal{I}$ 
for AOs on atom $A$, and $S^{A}_{\mu \nu}$ is the overlap matrix of these AOs.
The $\sum\limits^{\mathrm{on} \, A}_{\mu \nu}C^{A}_{\mu \mathcal I} S^{A}_{\mu \nu} C^{A}_{\nu \mathcal I}$ 
part is  thus the Mulliken charge of (the truncated) LMO $\mathcal{I}$ on atom A. 
Therefore, measure $M_{\mathcal{IJ}}$ sums the products of LMO $\mathcal{I}$ and 
LMO $\mathcal{J}$ Mulliken charges for all atoms, hence it can be interpreted as a discretized measure of the 
overlap of the two LMOs.
Orbital pairs for which $M_{\mathcal {IJ}}$ is large are more likely to be localized on the same or nearby atoms,
 and consequently may strongly interact.
The strong pair list is therefore extended with those pairs which exhibit a
considerable $M_{\mathcal{IJ}}$ measure. 
In practice, it is preferable to include the pairs which are indicated to be more strongly interacting 
 compared to their pair correlation energy. 
The selection of the additional strong pairs is thus controlled by the $h_{\mathrm{w}}$ parameter as 
 $M_{\mathcal{IJ}} / \delta E^{\mathrm{MP2}}_{\mathcal{IJ}} > h_{\mathrm{w}}$ relative to the pair correlation energies. 
In practice, we found that $g_{\mathrm{w}}=5$ and $h_{\mathrm{w}}=50 $ E$_\text{h}^{-1}$   are reasonable choices also for particularly 
challenging cases, while the strong pair list extension  introduce (practically) no changes
 to our previous strong pair list definition for organic molecules. 


\subsection{Local MP2 energy in the extended domains}\label{sec:ed_mp2}

The MP2 correlation energy contribution of each LMO is evaluated in its ED analogously to our RO LMP2 method \cite{ROLocalMP2},
so we focus on the steps required in the EDs  for LNO-CC computations.
The occupied subspace of the ED consists of its central LMO  and the strong LMO pairs of the central LMO. 
The atoms of the ED are defined to be the union of the extensive BP atom lists of all LMOs in the ED.
These large  BP atom lists are constructed with a completeness criterion of $T_{\mathrm{EDo}}=0.9999$ by default.
Then, each LMO is projected onto the AOs of its respective BP atom list, ensuring at most $1-T_{\mathrm{EDo}}$ (that is, here below 0.01\%) projection error
for accurate computations with these projected LMOs in the ED.
The projected LMOs are reorthogonalized by a specific combination of the Gram--Schmidt and L\"owdin symmetric (GSL) orthogonalization algorithms \cite{Mayervectors,MayervectorsGen}.
In the Gram--Schmidt step, the central LMO and all SOMOs are projected out from the DOMOs of the ED
 to ensure that they are not changed until the LNO construction step 
(see Section \ref{sec:lno}). Subsequently, the projected DOMOs are L\"owdin orthogonalized and all 
occupied orbitals are semi-canonicalized. For the latter, 
the spin up and spin down Fock matrix blocks of the ED are diagonalized separately
 in the spin up and spin down occupied orbital bases.
Next, the virtual subspace of the ED is constructed
 from the restricted PAOs centered on the atoms of the PAO center domain (PCD) of the ED.
The PCD of an ED is defined as the union of the compact BP atom lists of the LMOs in the ED, 
constructed with $T_{\mathrm{o}}=0.985$.
The selected PAOs are subsequently projected onto the AO basis of the ED.
To span the  spin down virtual subspace of the ED, the SO LMOs of the ED are 
appended to the spin down PAO list of the ED.
Finally, the occupied subspace is projected out from the virtual space of the ED, 
the virtual orbitals are L\"owdin canonical orthogonalized among themselves, and are 
semi-canonicalized in a spin unrestricted manner.

The ERIs of the ED are transformed to the MO bases of the ED using the DF approximation \cite{BoysShavittH3,DensFit}
and highly-optimized AO integral\cite{ERI} and local integral-transformation algorithms.\cite{LocalMP2,ROLocalMP2}
As demonstrated previously \cite{LocalMP2},
the set of auxiliary basis functions of the DF approximation can also be restricted to include those residing on atoms of the PCD.
The three-center DF integrals 
$(\tilde \mu \tilde \nu | X )$ 
are therefore
 computed only for a significantly restricted AO list ($\tilde \mu$) of the ED and for auxiliary functions ($X$) in the PCD.
The transformation to the MO bases of the ED is started with
an intermediate transformation step to the restricted occupied LMO basis of the ED. 
Since this first step is the most demanding, the overall cost of the open-shell DF integral transformation is
kept comparable to that of our closed-shell LMP2 algorithm.\cite{ROLocalMP2}

The four-center antisymmetrized ERIs are then only assembled for the single central LMO, e.g., a single $i^\prime$ index value, 
retaining the formally at most fourth-power-scaling operational complexity in the ED in terms of the dimensions of the ED bases:
\begin{equation}
\langle \tilde a \tilde b \Vert i^{\prime} \tilde j \rangle =
( \tilde a i^{\prime}  \vert \tilde b \tilde j ) - ( \tilde a \tilde j \vert \tilde b i^{\prime} ) =
K_{\tilde a i^{\prime} , \tilde b \tilde j} - K_{\tilde a \tilde j,\tilde b i^{\prime} } ,
\label{eqn:ed_eri}
\end{equation}
where the $\bf K$ tensors are computed from the DF integrals according to
\begin{equation}
\bf K = I V^{-1} I^{\mathrm{T}}  =
J I^{\mathrm{T}} \, .
\label{eqn:ed_df_int}
\end{equation}
In the above equation, $\bf I$ contains three-center DF integrals transformed to the ED MO basis [$(\tilde a \tilde j|X)=I_{\tilde a \tilde j, X}$], 
while $J_{\tilde a i^{\prime} ,Y} = \sum_{X}I_{\tilde a i^{\prime}, X} V^{-1}_{XY}$, and $V_{XY}=\left(X\vert Y\right)$ is the two-center auxiliary integral, whose inverse is computed using Cholesky-decomposition: ${\bf V}^{-1}= \left({\bf L L}^{\mathrm{T}}\right)^{-1}=\left({\bf L}^{-1}\right)^{\mathrm{T}}{\bf L}^{-1}$. 

Similar to the four-center  ERIs, the assembly of the MP1 amplitudes is only required for the fixed $i^\prime$ or $I^\prime$ values,
therefore, it is advantageous to compute them using either Cholesky-decomposition  or Laplace-transform. 
This way, the amplitudes can be written in closed form also in the non-canonical basis used here
 to avoid the redundant computation for other occupied index values \cite{LocalMP2,ROLocalMP2}.
Accordingly, the MP1 amplitudes of the ED are computed as
\begin{equation}
    t^{\tilde a \tilde B [1]}_{i^\prime \tilde J} = \sum_{\omega}\sum_{X}\bar I^{\omega}_{\tilde a i^\prime, X} \bar J^{\omega}_{\tilde B \tilde J, X} \, ,
\label{eqn:cholesky_amplitude}
\end{equation}
where the index $\omega$ runs over the Cholesky-vectors or the integration quadrature of the Laplace-transform.
The bar over integrals $I$ and $J$ denotes that these have been multiplied with the factor corresponding to the factorized
energy denominator. For example, $\bar J^{\omega}_{\tilde B \tilde J, X}=J_{\tilde B \tilde J, X}c^{\omega}_{\tilde B \tilde J}$, where $c^{\omega}_{\tilde B \tilde J}$ is either a Cholesky-vector element or if Laplace-transform  is used,
\begin{equation}
    c^{\omega}_{\tilde B \tilde J} = \sqrt{w_{\omega}}\exp \left(-\left(\varepsilon_{\tilde B} - \varepsilon_{\tilde J}\right)t_{\omega}\right)
\label{eqn:lt_coefficient}
\end{equation}
with $t_{\omega}$ and $w_{\omega}$ as a quadrature point and the
corresponding weight, respectively.
It is important to note that, for example, the Laplace-quadrature is also determined 
in a spin independent manner\cite{ROLocalMP2} enabling 
the assembly also of the mixed spin MP1 amplitudes, such as in eq \ref{eqn:cholesky_amplitude}.
Building the four-center integrals and MP1 amplitudes for all required spin cases leads to the 
ED MP2 energy contribution of the central LMO as
\begin{equation}
\begin{split}
\delta E^{\mathrm{MP2}}_{\mathcal{I}}\left(\mathcal{E}_{\mathcal{I}}\right) = \delta E^{\mathrm{MP2}}_{i^{\prime} }\left(\mathcal{E}_{\mathcal{I}}\right) + \delta E^{\mathrm{MP2}}_{I^{\prime} }\left(\mathcal{E}_{\mathcal{I}}\right) = \\
\sum_{\tilde a}t^{\tilde a \left[1\right]}_{i^{\prime} } f^{\tilde a}_{i^{\prime} }+\frac{1}{2}\sum_{\tilde a < \tilde b, \tilde j} t^{\tilde a \tilde b \left[1\right]}_{i^{\prime} \tilde j} \langle \tilde a \tilde b \Vert i^{\prime} \tilde
j \rangle + \frac{1}{2}\sum_{\tilde a \tilde B\tilde J}t^{\tilde a \tilde B \left[1\right]}_{i^{\prime} \tilde J}\langle \tilde a \tilde B | i^{\prime} \tilde J \rangle + \\
\sum_{\tilde A}t^{\tilde A \left[1\right]}_{I^{\prime} }f^{\tilde A}_{I^{\prime} }+ \frac{1}{2}\sum_{\tilde A < \tilde B, \tilde J}t^{\tilde A \tilde B \left[1\right]}_{I^{\prime} \tilde J}\langle \tilde A \tilde B \Vert I^{\prime} \tilde J
\rangle + \frac{1}{2}\sum_{\tilde A \tilde b, \tilde j}t^{\tilde A \tilde b \left[1\right]}_{I^{\prime} \tilde j}\langle \tilde A \tilde b | I^{\prime} \tilde j \rangle \, .
\end{split}
\label{eqn:cim_mp2_ed_energy}
\end{equation}

Here, it is important to note how the contributions from single excitations to the ED MP2 energy 
originate from two sources. First, as we employ the restricted open-shell determinant,
the unrestricted Fock matrices are not self-consistent and thus have off-diagonal elements even in the 
unrestricted canonical MO basis of the entire molecule. Second, as a consequence 
of truncating the occupied and virtual MOs of the ED, a small portion of the exact virtual 
space is mixed into the occupied ED MOs and vice versa. The latter contribution to the singles MP1 
amplitudes and hence to the MP2 energy of the ED is discarded in our closed-shell LMP2 variant, 
which, however, would undesirably discard the first contribution as well in the open-shell case. 
Since the first contribution originates from the off-diagonal block of the MO Fock 
matrix, while the second one appears mostly in the diagonal part of the MO Fock, the two
parts can be separated.\cite{ROLocalMP2} For that purpose, we also need to separate the 
off-diagonal Fock matrix blocks in the AO basis as
\begin{equation}
{\bf F^{\mathrm{OD}} }= {\bf F} - {\bf C}{\boldsymbol \epsilon}{\bf C}^\mathrm{T} \, ,
\label{eqn:off_diagonal_fock}
\end{equation}
where ${\bf F}^{\mathrm{OD}}$ and ${\bf F}$ are the off-diagonal and the full Fock matrices 
in the AO basis, respectively, while  ${\bf C}$ is the matrix of the unrestricted MO coefficients, 
and  ${\boldsymbol \epsilon}$ is a diagonal matrix with the corresponding orbital energies.
Then, we use ${\bf F}^\mathrm{OD}$  to evaluate off-diagonal Fock matrix elements in the MP2 energy of the ED in eq \ref{eqn:cim_mp2_ed_energy}. 
This ${\bf F}^\mathrm{OD}$ matrix vanishes if the reference orbitals of the molecules are exact eigenfunctions of the full Fock matrix,
such as in the case of our closed-shell methods where self-consistent HF reference is employed. 

Finally, we also note that the complete local MP2 energy can be constructed as a side product of 
an LNO-CC computation as
\begin{equation}
E^{\mathrm{LMP2}} = \sum_{\mathcal{I}}   \delta E^{\mathrm{MP2}}_{\scriptscriptstyle \mathcal{I}} \left(\mathcal{E}_{\scriptscriptstyle \mathcal{I}}\right)
+\sum_{\mathcal{I} < \mathcal{J}}^{\mathrm{distant}} \delta E^{\mathrm{MP2}}_{\scriptscriptstyle \mathcal{IJ}}\left(\mathcal{P}_{\scriptscriptstyle \mathcal{IJ}}\right) \, .
\end{equation}
This is beneficial as this local MP2 energy can be used, for example, for various composite energy 
expressions, such as local MP2 level basis set corrections. 


\subsection{Local natural orbitals}\label{sec:lno}
Due to the extensive cost of CCSD(T), the orbital spaces of the EDs should be further decreased after the 
MP2 part of the ED computation is completed. To that end, local natural orbitals (LNOs) are constructed 
as the eigenvectors of the second-order density matrix blocks of the ED built from
the MP1 amplitudes of eq \ref{eqn:cholesky_amplitude}.
The so obtained LNO list is truncated by retaining the most important occupied and virtual LNOs for the domain correlation energy contribution
  with occupation numbers above a threshold.

For this purpose, first, the occupied-occupied density matrix block contribution of central LMO $\mathcal I $
 is obtained in the semi-canonical ED basis for both spin cases as
\begin{equation}
\begin{split}
D^{\mathcal I}_{\tilde j \tilde k} =
\sum_{\tilde a < \tilde b}{t^{\tilde a \tilde b \left[1\right]}_{i^\prime \tilde j} t^{\tilde a \tilde b \left[1\right]}_{i^\prime \tilde k}}+
\sum_{\tilde A \tilde b}{t^{\tilde A \tilde b \left[1\right]}_{I^\prime \tilde j} t^{\tilde A \tilde b \left[1\right]}_{I^\prime \tilde k}} \, ,
\\
D^{\mathcal I}_{\tilde J \tilde K} =
\sum_{\tilde A < \tilde B}{t^{\tilde A \tilde B \left[1\right]}_{I^\prime \tilde J} t^{\tilde A \tilde B \left[1\right]}_{I^\prime \tilde K}}+
\sum_{\tilde a \tilde B}{t^{\tilde a \tilde B \left[1\right]}_{i^\prime \tilde J} t^{\tilde a \tilde B \left[1\right]}_{i^\prime \tilde K}} \, .
\end{split}
\label{eqn:occ_can_density}
\end{equation}
In order to reduce the computational cost of the integral transformation step (see Section \ref{inttrf}), it is beneficial 
to first construct an intermediate spin-restricted orbital  set also for the LNOs. 
To obtain spin-restricted LNOs, we transform the spin up and spin down density matrix blocks of eq \ref{eqn:occ_can_density}
to the restricted occupied basis of the ED and add together the resulting terms:
\begin{equation}
D^{\mathcal I}_{\hat l \hat m} =
\sum_{\tilde j \tilde k}C_{\hat l \tilde j} D^{\mathcal I}_{\tilde j \tilde k} C_{\hat m \tilde k}+ \sum_{\tilde J \tilde K}C_{\hat l \tilde J} D^{\mathcal I}_{\tilde J \tilde K} C_{\hat m \tilde K} \, .
\label{eqn:occ_rest_density}
\end{equation}
Here, $C_{\hat l \tilde j}$ ($C_{\hat l \tilde J}$) contains the spin up (spin down) semi-canonical
 to restricted LMO transformation coefficients of the ED.

Note that the restricted occupied basis contains both the central LMO and the SOMOs without mixing them with the rest of the 
truncated LMOs of the ED. This is the intentional result of using the GSL orthogonalization as described in Section \ref{sec:ed_mp2}
since the central LMO has to be kept exactly in the LIS, and we also want to keep the DO and SO
subspaces separated after the truncation of the LNO basis. 
Therefore, before diagonalizing the restricted  density matrix, its  $D^{\mathcal I}_{\mathcal I \hat m}$ and $D^{\mathcal I}_{\hat l \mathcal I}$ 
elements are replaced with zeros to avoid mixing the central LMO with the rest of the occupied ED orbitals upon diagonalization.
If the ED contains any SOMOs, the SOMO-DOMO blocks of the density matrix are also overwritten with zeros
in order to obtain a single set of restricted occupied LNOs by compressing only the DO subspace of the ED.
After diagonalizing this modified density matrix, 
the (restricted) occupied basis of the LIS ($\{\bar i\}$)  consists of the central LMO, all SOMOs, 
and those doubly occupied LNOs that exhibit occupation numbers higher than the occupied LNO threshold, $\varepsilon_\mathrm{o}$.
Then, the retained orbitals are semi-canonicalized using the spin up and spin down Fock matrices,
 yielding the final, unrestricted semi-canonical occupied LNO basis ($\{\underline i\}, \{\underline I\} $) in which the CCSD(T) contribution of the LIS is evaluated.

Notice that in domains where the central LMO is singly occupied, all MP1 amplitudes 
with an occupied index of $I^\prime$ are zero because there is no spin down electron on such orbitals.
Thus, for SO central LMOs, two of the four terms of eq \ref{eqn:occ_can_density} also vanish, 
which approximately  halves the number of non-zero density matrix terms compared to the case of DO central LMOs.
To treat the LNO construction with both SO and DO central LMOs on the same footing, 
we scale the density matrix corresponding to SO central LMOs by a factor of two. 
With this setting, we have found that the same $\varepsilon_\mathrm{o}$ occupied LNO threshold provides 
balanced accuracy for both domain types. 

Having the occupied LNO space at hand, we recommended to take advantage of the fact that 
the virtual LNOs only need to describe the correlation energy contribution of the retained occupied LNOs.\cite{LocalCC3}
To achieve this, 
the occupied indices of the MP1 amplitudes in the ED 
are transformed to the retained LNO basis before the virtual density matrix construction. This 
decreases the number of terms contributing to the  virtual-virtual density matrix elements and thus leads 
to a smaller number of virtual LNO occupation numbers being above the
corresponding threshold, $\varepsilon_\mathrm{v}$.\cite{LocalCC3}

It is worth noting a difference in the definition of the density matrix contribution of the central LMO to its
occupied (eq \ref{eqn:occ_can_density}) and virtual (eq \ref{eqn:virt_can_density} below) blocks. 
Namely, for the occupied-occupied block only a single occupied index can be selected to be the central LMO as the 
other two indices ($\tilde j$ and $\tilde k$ of eq \ref{eqn:occ_can_density}) are the indices of the density matrix block.
In contrast to that, virtual indices ($\tilde a$ and $\tilde b$ of eq \ref{eqn:virt_can_density} below) 
label the elements of the virtual-virtual density matrix block, thus 
there is some freedom of choice regarding which occupied index is set to be the central LMO 
and at what point of the derivation should one introduce the restriction for the central LMO index. 
This leads to three  different local virtual density matrix fragment variants, 
for which we provide derivations and further analysis in the Appendix. 
Out of the three choices the definition with the most balanced contribution for all spin cases was selected:
\begin{equation}
\begin{split}
D^{\mathcal I}_{\tilde a \tilde b}=
\frac{1}{2}\left(\sum_{\tilde c \underline j}{t^{\tilde a \tilde c \left[1\right]}_{i^\prime \underline j} t^{\tilde b \tilde c \left[1\right]}_{i^\prime \underline j}}+\sum_{\tilde C \underline J}
{t^{\tilde a \tilde C \left[1\right]}_{i^\prime \underline J} t^{\tilde b \tilde C \left[1\right]}_{i^\prime \underline J}}+\sum_{\tilde C \underline j}{t^{\tilde a \tilde C \left[1\right]}_{I^\prime \underline j}t^{\tilde b \tilde C \left[1\right]}_{I^\prime \underline j}}\right) \, , \\
D^{\mathcal I}_{\tilde A \tilde B}=
\frac{1}{2}\left(\sum_{\tilde C \underline J}{t^{\tilde A \tilde C \left[1\right]}_{I^\prime \underline J} t^{\tilde B \tilde C \left[1\right]}_{I^\prime \underline J}}+\sum_{\tilde c \underline j}
{t^{\tilde A \tilde c \left[1\right]}_{I^\prime \underline j} t^{\tilde B \tilde c \left[1\right]}_{I^\prime \underline j}}+\sum_{\tilde c \underline J}{t^{\tilde A \tilde c \left[1\right]}_{i^\prime \underline J}t^{\tilde B \tilde c \left[1\right]}_{i^\prime \underline J}}\right) \, .
\label{eqn:virt_can_density}
\end{split}
\end{equation}
Let us note that for closed-shell systems this virtual density matrix expression 
does not completely match the density matrix in our closed-shell LNO-CC method \cite{LocalCC,LocalCC2,LocalCC3,LocalCC4}
as our previous definition reduces to one of the other two expressions derived in the Appendix. 
Therefore the virtual density matrix expression of the closed-shell algorithm is updated to match the open-shell form defined here in the limit of applying the open-shell code to closed-shell molecules. This introduces a very small change to the closed-shell LNO-CCSD(T) energies much below the uncertainties corresponding to the LNO approximations.

In order to define an intermediate restricted LNO basis, 
the virtual density matrices of eq \ref{eqn:virt_can_density} built
 in the canonical ED basis are transformed to the restricted PAO basis of the ED. 
Then, the spin up and spin down components are added together analogously to the occupied case in eq \ref{eqn:occ_rest_density}.
Similar to the occupied density matrix fragment, 
half of the terms of eq \ref{eqn:virt_can_density} vanish in domains with a SO central LMO.
Therefore, in these domains the virtual density matrix  fragment is also scaled by a factor of two.
If  SOMOs are present in the ED, the SOMO-unoccupied block of the restricted density matrix is replaced with zeros, 
enabling the computation of  spin-restricted virtual LNOs upon the diagonalization of the density matrix modified in this way.
The restricted virtual basis of the LIS ($\{\bar a\}$) then consists of 
  the SOMOs and the retained virtual LNOs with occupation numbers larger than the threshold $\varepsilon_{\mathrm{v}}$.
The resulting restricted virtual LNO basis is then semi-canonicalized separately with the spin up and spin down Fock matrices to obtain 
the  unrestricted, semi-canonical virtual LNO basis of the LIS ($\{\underline a\}, \{\underline A\} $).


\subsection{Two-external integral transformations} \label{inttrf}

For the CCSD(T) computations in the LNO basis of the LIS,
the integral lists $( \bar i \bar j \vert X ) $, $( \bar a \bar i \vert X ) $, and $( \bar a \bar b \vert X ) $ are required.
By utilizing intermediate restricted LNO bases, this demanding step can be performed at a cost 
very close to the that of our closed-shell LNO-CC method. 
The $( \bar a \bar b \vert X ) $ two-external integral list poses a larger computational challenge
 due to the fact that a typical LIS contains about 4--5 times as many or more virtual LNOs as occupied ones.
To overcome a potential bottleneck for large systems corresponding to 
naive AO ($\to$ ED PAO) $\to$ LNO transformation algorithms, for our closed-shell LNO-CC approaches,
we implemented a significantly more efficient solution introducing 
approximate intermediate basis sets denoted as PAO' and LNO'.\cite{LocalCC3}
That  approach is generalized here to the open-shell case focusing on the algorithmic details, 
while we refer to our original discussion regarding the theoretical introduction and justification of the PAO' and LNO' functions.\cite{LocalCC3}

Briefly, an intermediate PAO' basis is introduced in each ED designed to provide 
a much more compact expansion of the virtual LNOs than the AO basis of the ED. For that purpose,
 we  project out the restricted DOMOs and SOMOs of the ED 
from the AOs of its PCD ($\mu^{\mathrm{PCD}}$):
\begin{equation}
\vert \mu^\prime \rangle = \left(1 - \sum_{\hat j}^{\mathrm{ED}}\vert \hat j \rangle \langle \hat j \vert \right)\vert \mu^{\mathrm{PCD}}\rangle =
\vert \mu^{\mathrm{PCD}}\rangle - \sum_{\hat j}^{\mathrm{ED}} O_{\hat j \mu^{\mathrm{PCD} }} \vert \hat j \rangle \, ,
\label{eqn:pao_prime}
\end{equation}
where $O_{\hat j \mu^{\mathrm{PCD} }}$ denotes the overlap integral $\langle \hat j \vert \mu^{\mathrm{PCD}} \rangle $. 
The resulting PAO' orbitals are not identical to the PAOs of the ED
 but the difference is small for our purposes and well controlled by the 
strong pair energy threshold ($\varepsilon_{w}$) and the ED occupied BP completeness criterion ($T_{ \mathrm{EDo}}$)
as discussed in detail for the closed-shell case.\cite{LocalCC3}
Therefore, the compact expansion in the PAO' basis instead of the AO basis of the ED can also be written for approximated virtual LNOs as
\begin{equation}
\vert \bar a^\prime \rangle = \sum_{\mu^\prime} A_{\mu^\prime \bar a^\prime}\vert \mu^\prime \rangle =
\sum_{\mu^{\mathrm{PCD} }}^{\mathrm{PCD}}A_{\mu^{\mathrm{PCD}} \bar a^\prime}\vert \mu^{\mathrm{PCD}}\rangle - \sum_{\hat j}^{\mathrm{ED}} B_{\hat j \bar a^\prime} \vert \hat j \rangle \, .
\label{eqn:lno_prime}
\end{equation}
Here, $\bf A$ collects the orbital coefficients of the  LNO' functions in the PAO' basis, while $\bf B = OA$.
Then, the corresponding two-external integrals can be written in the LNO' basis as
\begin{equation}
\begin{split}
    ( \bar a^\prime \bar b^\prime \vert X ) =&
\sum_{\mu, \nu}^{\mathrm{PCD}}A_{\mu \bar a^\prime}A_{\nu \bar b^\prime}( \mu \nu \vert X )
- \sum_{\nu, \hat j}^{\mathrm{PCD, ED}}B_{\hat j \bar a^\prime} A_{\nu \bar b^\prime}( \hat j \nu \vert X ) \\
&- \sum_{\mu,\hat k}^{\mathrm{PCD, ED}}A_{\mu \bar a^\prime}B_{\hat k \bar b^\prime}( \mu \hat k \vert X )
+ \sum_{\hat j,\hat k}^{\mathrm{ED}}B_{\hat j \bar a^\prime} B_{\hat k \bar b^\prime} ( \hat j \hat k \vert X ) \, .
\end{split}
\label{eqn:two_externals}
\end{equation}
The main benefit is that the summations over the AOs in the above equation run over the PCD, 
which is usually about 2-3 times smaller than the (complete) AO basis of the ED spanning the original virtual LNOs. 
We showed previously that this approach can lead to about a factor of  8-9 speedup 
 for large systems with saturated ED sizes with negligible  
 difference compared to the use of the original LNO basis. 

Compared to the two-external integrals, the $( \bar i \bar j \vert X ) $ and $( \bar i \bar b ^\prime \vert X ) $ integrals are 
obtained with relatively low computational costs by transforming, e.g.,
 the half-transformed integrals in the ED LMO basis [$( \mathcal{I}  \bar \mu \vert X ) $] to the LIS LNOs.
To complete the two-external integral lists, one additionally needs to evaluate the SOMO-virtual and SOMO-SOMO blocks of 
$( \bar a^\prime \bar b^\prime \vert X ) $, which will contribute to the two-external integrals in the spin down unrestricted basis. 
However, since we so far worked in a restricted basis, where the SOMOs play the dual role of occupied spin up and virtual spin down orbitals,
 the required blocks are equivalent to the SOMO-virtual and the SOMO-SOMO block of the spin up $( \bar i \bar b^\prime \vert X ) $ and $( \bar i \bar j \vert X )$ integrals, respectively.
The complete set of restricted two-external integrals of the LIS can therefore be efficiently compiled from the three sets of integrals obtained so far. 
As the last step of the two-external integral transformations, the restricted $(\bar a^\prime \bar b^\prime \vert X)$ integrals are transformed to the semi-canonical spin up and spin down bases of the LIS, which represents a negligible additional cost compared to the closed-shell algorithm.
For the sake of simplifying the notation, the prime distinction of the virtual LNOs will be omitted in the remaining sections.

\subsection{Natural auxiliary functions}\label{sec:naf}

Before utilizing the above  three-center integrals to assemble the four-center ERIs of the LIS, 
their auxiliary function index is compressed using the natural auxiliary function (NAF) technique.\cite{NAF,FNONAF}
NAFs are constructed to find the best low-rank approximation of
 the $J_{\underline p \underline q, X}$ and $J_{\underline P \underline Q, X}$ three-center integrals of the LIS.
For efficiency considerations, instead of preforming the singular value decomposition of the 
three-center integral tensors, we can build the following $\bf W$ intermediates:
\begin{equation}
\begin{split}
W^{\uparrow}_{XY} = \sum_{\underline p \le \underline q} \left( \underline p \underline q \vert X \right) \left( \underline p \underline q \vert Y \right) \, , \, \, \, 
W^{\downarrow}_{XY} = \sum_{\underline P \le \underline Q} \left( \underline P \underline Q \vert X \right) \left( \underline P \underline Q \vert Y \right) \, .
\end{split}
\label{eqn:naf_w}
\end{equation}
It is important to note that for the assembly of four-center ERIs with mixed spin, the 
auxiliary basis should be the same for both the spin up and spin down DF integrals. 
Therefore, we diagonalize the spin averaged ${\bf W = (W^{\uparrow} + W^{\downarrow})} / 2$ matrix\cite{NAF,FNONAF} 
and retain its eigenvectors as NAFs with eigenvalues above $\varepsilon_{\mathrm{NAF}}$.
We also note that the NAFs obtained in this manner are equivalent to those of the closed-shell LNO-CCSD(T) algorithm in the closed-shell limit
 due to the factor of $\frac{1}{2}$ in the definition of  $\bf W$.

\subsection{Correlation energy calculation in the LIS} \label{sect:lisE}
Having the DF integrals transformed to the LNO and NAF bases of the LIS, one can proceed with 
writing the CCSD and (T) correlation energy contributions of  eqs \ref{eqn:cim_ccsd_energy} and \ref{eqn:cim_pt_energy} 
specifically for the LNO approximations in the LIS.

First, the CCSD correlation energy  contribution is evaluated in the LIS as
\begin{equation}
\begin{split}
\delta E^{\mathrm{CCSD}}_{\mathcal I}  =
&\sum_{\underline a} t^{\underline a}_{i^{\prime} }f^{\underline a}_{i^{\prime} }
+ \frac{1}{4} \sum_{\underline a\underline b\underline j}\tau^{\underline a\underline b}_{i^{\prime} \underline j}\langle \underline a\underline b \Vert i^{\prime} \underline j \rangle
+ \frac{1}{2}\sum_{\underline a\underline B\underline J}\tilde \tau^{\underline a\underline B}_{i^{\prime} \underline J}\langle \underline a\underline B | i^{\prime} \underline J \rangle \\
&+\sum_{\underline A} t^{\underline A}_{I^{\prime} }f^{\underline A}_{I^{\prime} }
+ \frac{1}{4} \sum_{\underline A\underline B\underline J}\tau^{\underline A\underline B}_{I^{\prime} \underline J}\langle \underline A\underline B \Vert I^{\prime} \underline J \rangle
+ \frac{1}{2}\sum_{\underline A\underline b\underline j}\tilde \tau^{\underline A\underline b}_{I^{\prime} \underline j}\langle \underline A\underline b | I^{\prime} \underline j \rangle \, .
\end{split}
\end{equation}
The CCSD  amplitudes are obtained by solving the CCSD equations in the basis of the LIS,
 relying on our hand-coded, highly-optimized, and semi-integral direct unrestricted CCSD(T) implementation in \textsc{Mrcc}\cite{MRCC,mrcceng3}.
The computational cost of a CCSD iteration scales in total as the sixth power of the occupied ($\underline n_{\mathrm{o}}$)  and virtual ($\underline n_{\mathrm{v}}$)  LIS basis dimension [e.g., $\mathcal{O}(\underline n_{\mathrm{o}}^2 \underline n^{4}_{\mathrm{v}})$], which may 
take a significant portion of the total wall time of a LNO-CCSD(T) computation.
In comparison with the closed-shell alternative, the computational cost of the LIS CCSD part
 is roughly three times higher, stemming from the three times as many terms in the open-shell CCSD equations.

Regarding the (T) correction of the LIS, 
its naive implementation would scale with the seventh power of the LIS dimensions 
[i.e., $\mathcal{O}(\underline n^{3}_{\mathrm{o}} \underline n^{4}_{\mathrm{v}})$].
This can be reduced to sixth-power scaling in the LIS if we recognize that 
we only need the triples amplitudes for a fixed central LMO index to evaluate 
the $\delta E^{\mathrm{(T)}}_{I}$ correlation energy contribution since the 
perturbative (T) triples amplitudes are not coupled, at least in a semi-canonical basis.
For that purpose, we utilize an  orbital invariant form of the (T) expressions using the Laplace-transform in our LNO methods.\cite{LaplaceT}
In this way, the triples energy denominators in the LIS, $D^{\underline{abc}}_{\underline{ijk}}$, are factorized as 
\begin{equation}
    \frac{1}{D^{\underline{abc}}_{\underline{ijk}}} = \int^{\infty }_{0} \mathrm{e}^{-D^{\underline{abc}}_{\underline{ijk}}s}\mathrm{d}s
    \approx \sum_{q}^{n_q} \omega_q \mathrm{e}^{-D^{\underline{abc}}_{\underline{ijk}} s_q} \, ,
\label{eqn:laplace_transform}
\end{equation}
utilizing $n_\mathrm{q}$ pieces of quadrature points $s_q$, with quadrature weights $\omega_q$.
Writing the exponential of the energy denominators as products of exponentials of single orbital energies,
 they can be absorbed in the intermediates contributing to the triple excitation amplitudes.
Then, the latter can be directly computed in a basis that contains the  central LMO,
enabling an algorithm with a much more favorable scaling in the LIS
 [i.e., $\mathcal{O}( n_{ q} \underline n^{2}_{\mathrm{o}}   \underline n^{4}_{\mathrm{v}})$].

Utilizing this Laplace-transform approach, introduced in detail in Ref.~\citenum{LaplaceT}, 
the (T) contribution of central LMO $\mathcal I$ is given as
\begin{equation}
\begin{split}
    \delta E^{\mathrm{(T)}}_{\mathcal{I}} = \frac{1}{3}\sum_{q}\biggl( &
\sum_{\substack{\underline{a} < \underline{b} < \underline{c} \\ \underline{j} < \underline{k} }} \bar t^{\underline{abc}}_{i^\prime\underline{jk},q}\bar W^{\underline{abc}}_{i^\prime\underline{jk},q} +
\sum_{\substack{\underline{a} \, \underline{B} < \underline{C} \\ \underline{J} < \underline{K} }} \bar t^{\underline{aBC}}_{i^\prime\underline{JK},q}\bar W^{\underline{aBC}}_{i^\prime\underline{JK},q} +
\sum_{\substack{\underline{a} < \underline{b} \, \underline{C} \\ \underline{jK} }} \bar t^{\underline{abC}}_{i^\prime\underline{jK},q}\bar W^{\underline{abC}}_{i^\prime\underline{jK},q} \\
        + &\sum_{\substack{\underline{A} < \underline{B} < \underline{C} \\ \underline{J} < \underline{K} }} \bar t^{\underline{ABC}}_{I^\prime\underline{JK},q}\bar W^{\underline{ABC}}_{I^\prime\underline{JK},q}
+ \sum_{\substack{\underline{A} \, \underline{b} < \underline{c} \\ \underline{j} < \underline{k} }} \bar t^{\underline{Abc}}_{I^\prime\underline{jk},q}\bar W^{\underline{Abc}}_{I^\prime\underline{jk},q}
+ \sum_{\substack{\underline{A} < \underline{B} \, \underline{c} \\ \underline{Jk} }} \bar t^{\underline{ABc}}_{I^\prime\underline{Jk},q}\bar W^{\underline{ABc}}_{I^\prime\underline{Jk},q}
\biggr) \, .
\end{split}
\label{eqn:local_pt}
\end{equation}
Here, the overbar of $\bar t$ and $\bar W$ indicates that these quantities contain the Laplace energy denominator 
factors absorbed as defined in Ref.~\citenum{LaplaceT}.
The formal scaling of evaluating eq \ref{eqn:local_pt} is similar to that of its closed-shell analogue \cite{LaplaceT}, with a larger prefactor of about 3 due to the increased number of spin cases. The number of quadrature points is determined system specifically and, matching our closed-shell experience \cite{LaplaceT}, the resulting 3(--4) points provide sufficiently high accuracy for the (T) contribution.

\subsection{Comparison with the closed-shell LNO-CCSD(T) method}\label{sec:closed_shell_comparison}

Here, we compare the computational requirements of
the most important steps of the restricted open-shell LNO-CCSD(T) algorithm to their closed-shell counterparts.
The computationally demanding part of the MP2 pair energy evaluation 
is the transformation and contraction of the multipole approximated integrals,
 of which there are four times as many in the open-shell case. 
However, even for the largest systems of 500--600 atoms considered here, 
the pair energy evaluation  takes only a few percent of the total run time.
The next important algorithmic step is the three-center integral transformation in the ED basis.
Even though there are twice as many integrals to transform as in the closed-shell case, 
all transformation steps of considerable demand are kept at the same cost as in the closed-shell case by utilizing the intermediate restricted ED basis.
Next, the MP1 amplitudes, the MP2 energy, and the density matrices in the ED are computed from the available ERIs of the ED.
As these steps have to be carried out in the unrestricted semi-canonical basis,
their operation count requirement is roughly 2--3 times as large as that of the closed-shell analogue.
As the next step, the ERIs necessary for the LIS CCSD(T) calculations are transformed to the LNO and NAF bases.
Again, all transformation steps in the LIS with considerable cost are performed in the intermediate restricted virtual LNO basis to keep the 
operation count close to that of the closed-shell case.
Finally, the correlation energy contribution of the central LMO is computed in the LIS using an unrestricted CCSD(T) 
formalism requiring about three times as many operations as closed-shell CCSD(T). Since the LIS CCSD(T) part usually takes 
about half of the total LNO-CCSD(T) wall time, and our unrestricted CCSD(T) implementation is less well parallelized 
than the restricted one\cite{MPICCSDpT}, one can expect at least twice as high costs for the open-shell LNO-CCSD(T) compared to its
closed-shell counterpart for molecules of similar wave function complexity.

\subsection{Approximate long-range spin polarization}
While for small to medium-sized systems, the additional expense of using unrestricted CCSD(T) formalism is affordable with the present as well as 
other local CCSD(T) methods\cite{OS_DLPNO-CCSD(T),PNO-UCCSD(T)},
the same might not be the case for demanding, large-scale local CCSD(T) computations. 
Recently, we introduced the idea of approximating the long-range spin polarization at a lower level of theory, such as our distant pair MP2 model, 
which turned out to be very effective in our local MP2 implementation.\cite{ROLocalMP2}

To introduce this approach also at the LNO-CCSD(T) level, let us consider an extended system with a limited amount of SOMOs, which 
are often located around a well-defined, narrow region of the molecule. 
We activate this long-range spin polarization approximation domain specifically
when the central LMO of a given domain does not couple strongly to any SOMOs, i.e., when 
 there are only DOMOs and no SOMO  within the ED (and LIS).
Thus, when this approach is activated, the small, spin-polarized correlation effects between the distant LMO-SOMO pairs are
taken into account in the final LNO-CCSD(T) correlation energy via the multipole approximated MP2 pair energies.
In other words, if there are no SOMOs in the EDs of the LMOs that form distant pairs with all SOMOs, then
the only spin-polarization effect to take into account in those EDs comes from the spin-dependence of the domain Fock matrices. 
This manifests in the slight splitting of the spin-up and spin-down LMOs and PAOs of the ED due to their semi-canonicalization with the corresponding spin-dependent Fock matrices.
However, since the central LMO of such EDs forms distant pairs with all SOMOs, the AOs around the SOMOs are mostly
not part of the ED. Then, one can expect that the blocks of the spin-dependent Fock matrices corresponding to AOs far away from the SOMOs
are moderately spin-polarized and 
that the magnitude of this secondary spin-polarization effect in the ED energy contribution is small.
Therefore, if we approximate this spin polarization effect by replacing the 
spin dependent ED Fock matrix blocks with their spin-averaged counterpart, then all orbital sets of the ED and LIS become spin independent. 
Then, we can utilize the restricted local MP2 and LNO-CCSD(T) algorithms to evaluate the ED and LIS correlation energy contributions 
for the domains far away from the SOMOs. 
As a consequence, in large systems where the SOMOs are well localized to a single region of the molecule, 
we can take advantage of  the more efficient restricted formalism and thus approach 
the computational effectiveness of a completely closed-shell calculation. \cite{ROLocalMP2}


\subsection{Scaling of the algorithm}

The presented open-shell LNO-CCSD(T) method, just as its closed-shell analogue,
 achieves asymptotically linear scaling with the system size for its rate determining steps.
Since the size of the EDs saturates for sufficiently large systems, 
all computations performed in these domains exhibit asymptotically linear scaling.\cite{ROLocalMP2}
On the other hand, computations performed on the whole system, such as the occupied orbital localization, the PAO construction, and the pair energy computation, scale as the third, third, and second power of the system size, respectively.
However, the cumulative run time of these computations is negligible compared to the total run time, even for the largest systems considered here.
Moreover, especially for molecules of several hundreds of atoms, the SCF computation can potentially take a considerable portion of the run time. 
The implementation for the HF exchange term in {\sc Mrcc} is in principle also asymptotically linear scaling\cite{LocalMP2,Multipole}
(alongside an efficient cubic-scaling DF-based Coulomb term). However, 
the HF exchange computations reach the linear scaling system size range for much larger molecules than our
 local correlation methods, which is typically above a few thousand atoms for three 
dimensional structures. Below that size range, about cubic-scaling can be expected, 
as the local DF capability of our HF exchange implementation can effectively reduce the scaling already above a few hundred atoms.

In terms of memory requirements, the open-shell LNO-CCSD(T) algorithm requires the storage of six matrices with dimensions equal to the total number of AOs
of the entire molecule, while the preceding SCF procedure needs eight such matrices. These quadratic-scaling arrays are only relevant above tens of thousands of AOs.
Furthermore, the remaining  arrays allocated within  the EDs are asymptotically constant in size, and their memory requirement is 
also highly optimized following the ideas exploited in our restricted LNO-CCSD(T) code.\cite{LocalCC3,LocalCC4}  The shared-memory 
parallelization model used within a single node also contributes to the highly memory economic nature of our implementation. 


\section{Computational details and test systems} \label{sec:compdet}

\subsection{Technical details}
The open-shell LNO-CCSD(T) method as presented here is implemented in the development version of the \textsc{Mrcc} quantum chemical program suite\cite{mrcceng3,MRCC}
and will be made available in a forthcoming release of the \textsc{Mrcc} package.
The default or \texttt{Normal} truncation thresholds of the local approximations 
 and their corresponding keywords  are collected in Table \ref{tab:thresholds}.
These default values are equivalent to those utilized in the most recent closed-shell formulation of LNO-CCSD(T) \cite{LocalCC3,LocalCC4}.

In all HF and reference canonical CCSD(T) calculations the DF approximation was employed using the \textsc{Mrcc} program.
To accelerate the HF calculations for the largest systems (containing more than 500 atoms), 
the HF exchange contribution was evaluated in local fitting domains, as described in Section \ref{sec:scf}.
The core electrons (including the subvalence electrons of the iron and cobalt atoms) were kept frozen in all correlated computations.
The remaining occupied orbitals were localized with the algorithm of Boys \cite{BoyLoc} separately for the DO and SO subspace.

The calculations presented make use of the triple-$\zeta$ valence basis set including polarization functions (def2-TZVP) developed by Weigend and Ahlrichs \cite{AhlrichsDef2} and Dunning's (augmented) correlation-consistent polarized valence basis sets [(aug-)cc-pV$X$Z, $X$ = D, T and Q] \cite{PVXZ1stRow}, with the revised aug-cc-pV($X$+d)Z basis sets used for second-row atoms. \cite{PV(X+D)Z}
Additionally,
the triple-$\zeta$, weighted core-valence  basis of Balabanov and Peterson (cc-pwCVTZ) was utilized for the iron atom of the Fe(III) complex investigated in Section \ref{sec:lno_selection},  following the work of  Rado\'n. \cite{radon_iron_complexes} 
The appropriate auxiliary basis sets of Weigend et al. were used for all AO bases. \cite{WeigendDFBas}.
The extrapolations of the HF \cite{KMSCFCBS} and the correlation energies \cite{CorrConv} towards the
complete basis
set (CBS) limit were performed according to the standard expressions.

The relative energy deviations with respect to some reference energy ($E_{\mathrm{ref}}$) are obtained as $(100\%) \cdot (E_{\mathrm{LNO-CCSD(T)}} - E_\mathrm{ref}) / E_\mathrm{ref}$.
To characterize the achieved accuracy on various test sets the following statistical measures were utilized: the maximum absolute error (MAX), the mean absolute error (MAE), and, to measure the consistency of errors, the standard deviation of the absolute errors (STD).
The presented timing measurements were performed using 2.5 GHz Intel Xeon Gold 6148 processors
 with 20 physical cores and  at most 192 GB of total memory in dual-socket nodes.
\begin{table}[h]
    \centering
    \begin{tabular}{ccc}
         Symbol & Keyword\textsuperscript{\emph{a}} & Value \\\hline
         $\varepsilon_\mathrm{o}$ & \texttt{lnoepso} & $10^{-5}$ \\
         $\varepsilon_\mathrm{v}$ & \texttt{lnoepsv} & $10^{-6}$ \\
         $\varepsilon_\mathrm{w}$ [E\textsubscript{h}] & \texttt{wpairtol} & $10^{-5}$ \\
         $T_\mathrm{EDo}$ & \texttt{bpedo} & 0.9999 \\
         $\varepsilon_\mathrm{NAF}$ [E\textsubscript{h}] & \texttt{naf\_cor} & $10^{-2}$ \\
         $T_\mathrm{LT}$ & \texttt{laptol} & $10^{-2}$ \\
         $g_\mathrm{w}$ & \texttt{epairscale} & 5 \\
         $h_\mathrm{w}$ [E$_\text{h}^{-1}$]  & \texttt{epairestfact} & 50  
    \end{tabular}
    \caption{Default threshold values used in this study.}
    {\textsuperscript{\emph{a}} Name of the corresponding keyword in the {\sc Mrcc} program package.\cite{mrcceng3,MRCC}}
    \label{tab:thresholds}
\end{table}

\subsection{Benchmark sets and test systems}
The accuracy of the open-shell LNO-CCSD(T) correlation energies and energy differences are benchmarked on three test sets containing small to medium sized molecules.
The first test set contains the radical stabilization energies (RSEs) of 30 small organic molecules (RSE30).\cite{liu_phd}
These structures were selected from the RSE43 compilation\cite{GMTKN30} and reoptimized in ref \citenum{PNO-UCCSD}.
Next, a test set of adiabatic ionization potentials for 21 organic molecules (IP21) is considered
as defined in ref \citenum{PNO-UCCSD} (with structures reoptimized in ref \citenum{ROLocalMP2}).
Lastly, singlet--triplet energy gaps of aryl carbenes (AC) are also investigated invoking the AC12 test set as compiled in ref \citenum{aryl_carbenes}.

\begin{figure}
    \centering
    \includegraphics[width=0.7\columnwidth]{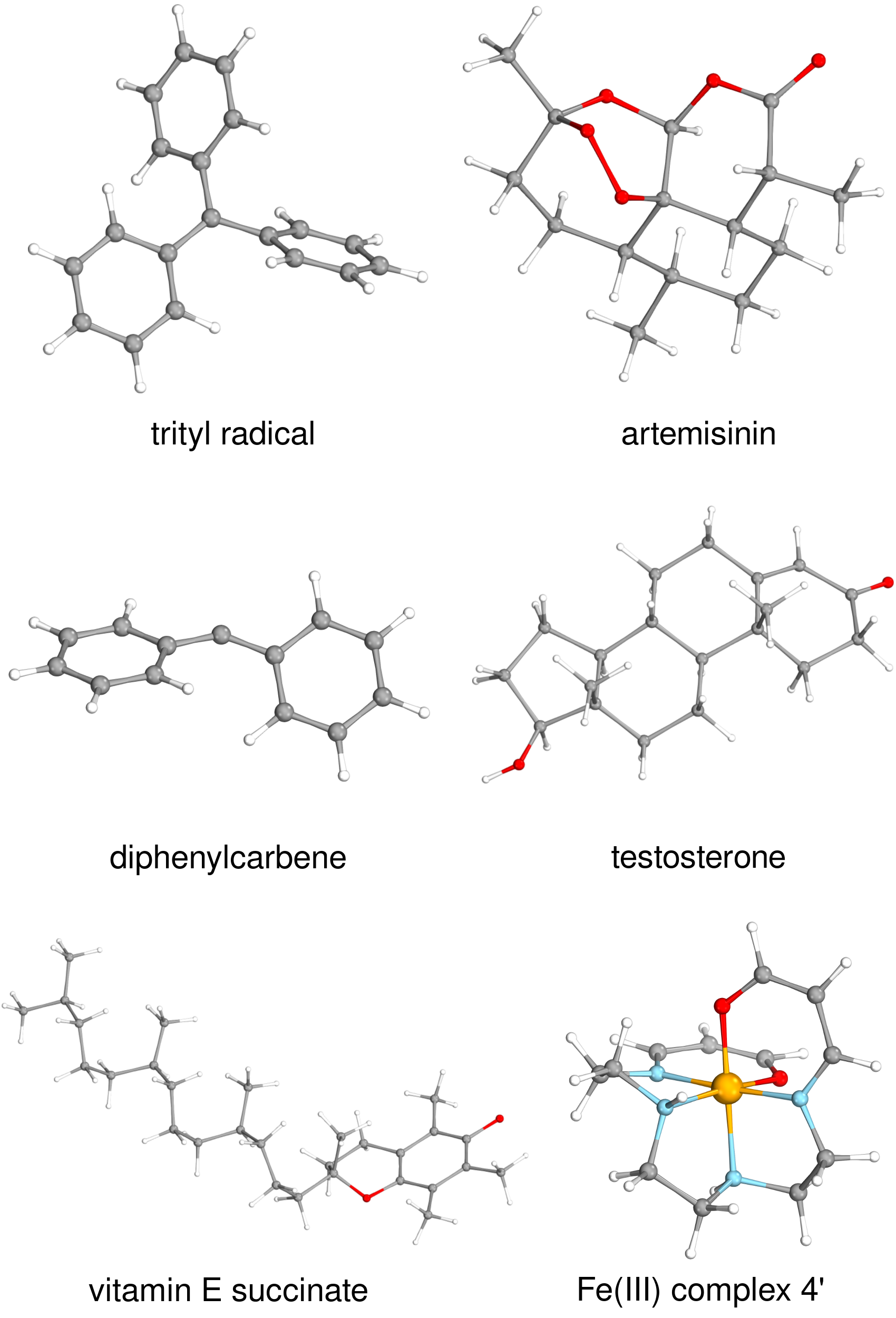}
    \caption{Structures of the six systems of intermediate size used to investigate the convergence of the local approximations. Top row: trytil radical and triplet artemisinin; middle row: diphenylcarbene in its triplet state and testosterone cation; bottom row: vitamin E succinate radical and the 4' Fe(III) complex of Rado\'n\cite{radon_iron_complexes} considered both with doublet and sextet multiplicities.}
    \label{fig:medium_structures}
\end{figure}

To investigate the convergence of the local approximations as a function of the corresponding thresholds,
 six larger open-shell systems containing 23--80 atoms are also selected.
The structures of vitamin E succinate, trityl radical, and artemisinin are obtained from ref \citenum{DLPNO_CCSD_OS}, testosterone is taken from ref \citenum{PNO-UCCSD}, whereas diphenylcarbene comes from the AC12 test set of ref \citenum{aryl_carbenes}.
Lastly, the 4' Fe(III) complex of ref \citenum{radon_iron_complexes} is also investigated, constructed by replacing the methyl groups of [Fe(acac\textsubscript{2}trien)]\textsuperscript{+} with hydrogens (where H\textsubscript{2}acac\textsubscript{2}trien is the Schiff base obtained from the 1:2 condensation of triethylenetetramine with acetylacetone).
The structures of these medium-sized systems are depicted in Figure \ref{fig:medium_structures}.

\begin{figure}
    \includegraphics[width=\textwidth]{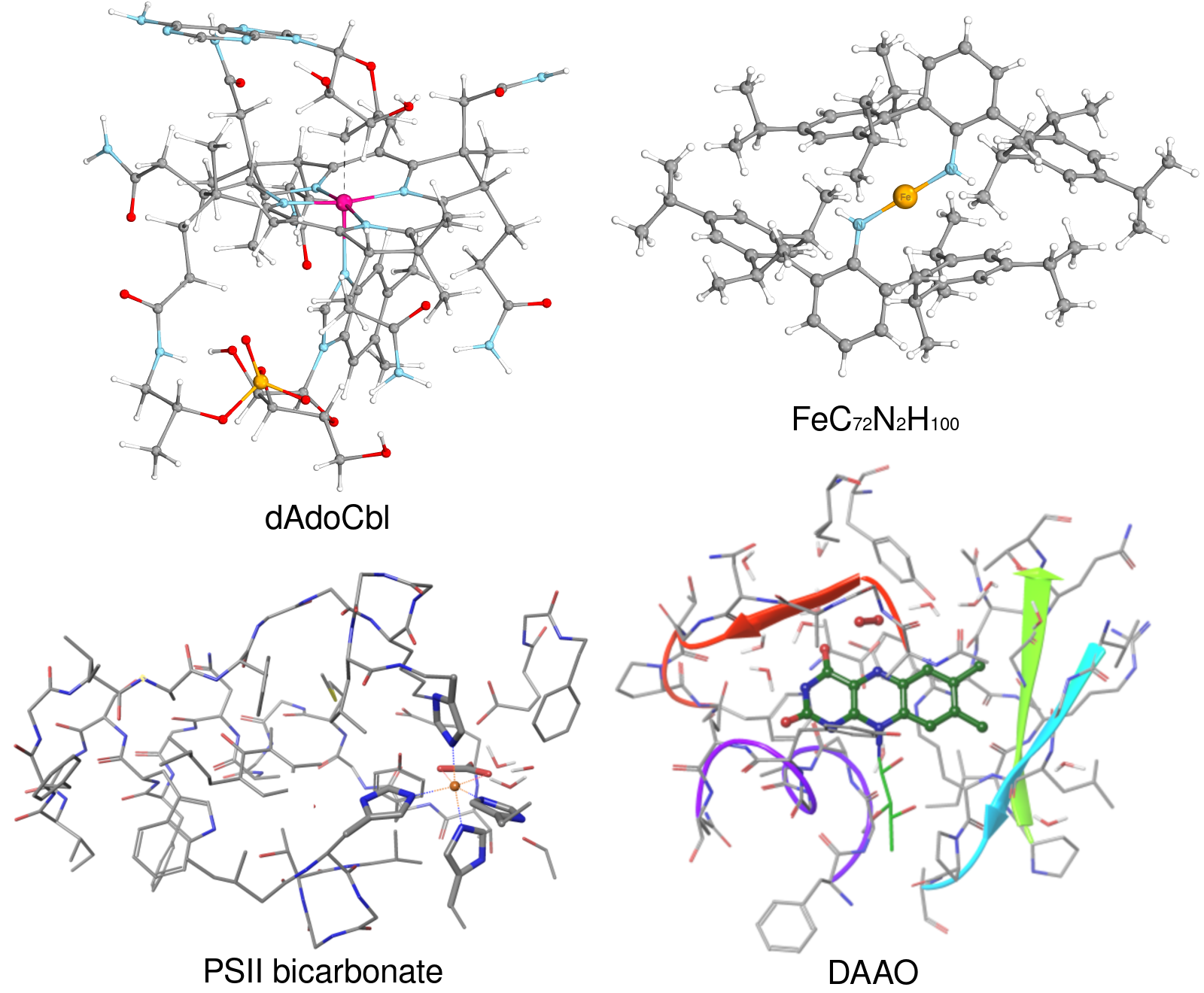}
    \caption{The structure of the four largest systems considered. Top row:  the 5'-deoxyadenosylcobalamin (209 atoms, out of which the Cob\textsuperscript{II}alamin (Cbl) radical part has 179 atoms) and the 175-atom iron(II) complex FeC\textsubscript{72}N\textsubscript{2}H\textsubscript{100}. Bottom row: a 565-atom model of photosystem II bicarbonate and the O1\textsuperscript{T} structure of \textsc{d}-amino-acid oxidase containing 601 atoms.}
    \label{fig:large_systems}
\end{figure}

To demonstrate the performance of the open-shell LNO-CCSD(T) method,  large-scale calculations were also carried out on four systems of 175--601 atoms (see Figure \ref{fig:large_systems}).
Among these is the 175-atom Fe(II) complex taken from ref \citenum{SparseMapsNEVPT2}, both in its triplet and quintet spin state.
Moreover, the Cob\textsuperscript{II}alamin (Cbl) radical containing 179 atoms, a product of the homolytic bond breaking of the coenzyme B\textsubscript{12} (5'-deoxyadenosylcobalamin, dAdoCbl), was also considered \cite{coenzyme_b12}.
The largest systems investigated are the models of photosystem II (PSII) bicarbonate \cite{DLPNO_CCSD_OS} containing 565 atoms and of the \textsc{d}-amino-acid oxidase (DAAO) \cite{DAAO_structure} made up of 601 atoms.
The singlet and triplet spin states of PSII bicarbonate were calculated utilizing the def2-TZVP basis set and QRO reference determinants.
The particular complexities and technicalities of this reference 
computation are discussed in ref \citenum{ROLocalMP2}.
In the case of DAAO, two steps along the oxidative half-reaction of \textsc{d}-alanine oxidation are taken from Kiss et al.\cite{DAAO_structure}, where the reduced flavin adenine dinucleotide (FAD) moiety of DAAO is oxidized by O$_2^-$ to produce the oxidized form of FAD and H\textsubscript{2}O\textsubscript{2}.
The corresponding reactant and product structures labeled by O1\textsuperscript{T} and O3\textsuperscript{CSS} in ref \citenum{DAAO_structure} are
also depicted in Figure 5 of ref \citenum{ROLocalMP2}.


\section{Convergence of the local approximations}\label{sec:convergence}
In this section, the effect of the local approximations of the open-shell LNO-CCSD(T) method on its accuracy is investigated individually.
To this end, the thresholds associated with such approximations are systematically varied, 
and the resulting energy deviations from approximation-free DF-CCSD(T) or other reference methods are observed.
Almost all of the employed approximations were extensively benchmarked in our related open-shell LMP2 and closed-shell LNO-CCSD(T) studies \cite{LocalMP2,LocaldRPA,LocalCC3,LocalCC4,ROLocalMP2}.
Consequently, the presented tests focus on the parameters with the largest impact on accuracy: the strong pair energy threshold $\varepsilon_\mathrm{w}$ and the LNO occupation limits $\varepsilon_\mathrm{o}$ and $\varepsilon_\mathrm{v}$.
Additionally, approximations that were previously not employed in the context of either 
any open-shell local CCSD(T) implementation or our open-shell LMP2 method, such as the distant spin polarization approximation discussed in Section \ref{sec:closed_shell_comparison} and the NAFs introduced in Section \ref{sec:naf}, are also investigated.
For the remaining truncation thresholds, affecting closed and open-shell systems similarly, such as the BP parameters of the LMO atom list definition, the previously benchmarked values are adapted, which will be tested in combination with all approximations in Section \ref{sec:statistics}.

\subsection{LNO selection}\label{sec:lno_selection}

First, the errors introduced by the truncation of the LNO basis of the LISs are illustrated on three systems.
The system sizes were chosen to be as large as possible so that the LNO truncation is already active in all domains (with up to 50\% of the occupied and 85\% of virtual LNOs discarded), while still keeping the possibility of using DF-CCSD(T) energies as reference.
The one-particle basis set of cc-pVTZ is chosen for the description of the hydrogen addition of the triphenylmethyl radical, while cc-pVQZ is used for the calculation of the singlet-triplet gap of diphenylcarbene. 
A composite basis set composed of cc-pwCVTZ for the iron atom, cc-pVTZ for the atoms connected to the iron atom, and cc-pVDZ for all other atoms is chosen to investigate the doublet-sextet gap of the Fe(III) complex, following Rado\'n.\cite{radon_iron_complexes}
Since our previous studies found that setting the ratio of the virtual and occupied LNO selection thresholds ($\varepsilon_\mathrm{o} / \varepsilon_\mathrm{v}$) to 10 gives satisfactory results\cite{LocalCC,LocalCC2,LocalCC3,LocalCC4}, this ratio was kept in this work.

\begin{figure}
\centering
\includegraphics[width=\textwidth]{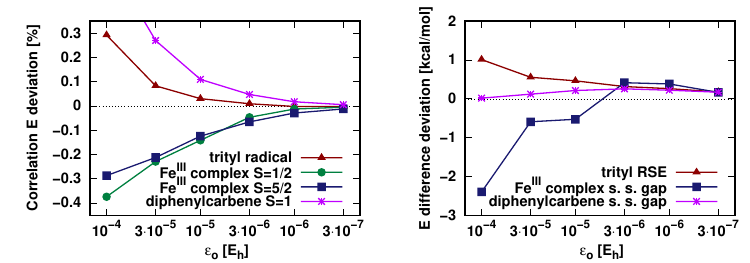}
\caption{Deviations of the LNO-CCSD(T) energies from reference DF-CCSD(T) values as a function of the LNO selection thresholds.
Left panel: relative correlation energy differences, right panel: deviations in energy differences (s. s. = spin state).
The ratio of the occupied and virtual LNO selection thresholds were kept constant, $\varepsilon_\mathrm{o}/\varepsilon_\mathrm{v}=10$.
For a description of the systems and basis sets utilized, see Section \ref{sec:lno_selection}.}
\label{fig:lno_conv}
\end{figure}

As can be seen from the left panel of Figure \ref{fig:lno_conv}, the correlation energy errors decrease monotonically and converge to
 zero when the LNO selection thresholds are tightened (the approximations other than the LNO truncation are turned off in these tests).
Moreover, the 99.9\% accuracy approached already with the default setting of $\varepsilon_\mathrm{o} = 10^{-5}$ and 
$\varepsilon_\mathrm{v} = 10^{-6}$ is highly satisfactory for such non-trivial systems.
Considering the right panel of the same figure, the three energy differences of the associated chemical processes also exhibit good convergence.
The singlet-triplet energy gap of  diphenylcarbene 
with almost perfect error cancellation already with the loosest thresholds
 is probably an exception, representative only for very local properties, like 
spin state differences, localized mostly on a few atoms.
Nonetheless, the deviations introduced by the LNO truncation for this system remain below 0.3 kcal/mol for all investigated $\varepsilon_\mathrm{o}$ values,
and the trityl RSE and the doublet-sextet gap of the Fe(III) complex also drop below the satisfactory 0.5~kcal/mol value after $\varepsilon_\mathrm{o}=10^{-5}$.
Thus, the default LNO threshold values of $\varepsilon_\mathrm{o}=10^{-5}$ and $\varepsilon_\mathrm{v}=10^{-6}$ are selected, which are
 recommended also based on our
closed-shell LNO-CCSD(T) benchmarks\cite{LocalCC3,LocalCC4} as both the correlation energy and energy difference deviations are sufficiently converged (with at most 0.14\% and 0.53 kcal/mol deviations, respectively, for the examples of this section). We note in passing that this value is not directly comparable to 
the frozen NO\cite{FNONAF,FNOExt} or pair NO\cite{OS_DLPNO-CCSD(T)} threshold settings 
as only the strong pair LMOs are included in our orbital specific LNO density matrices (Sect.~\ref{sec:lno}), 
while the summation is not restricted for the frozen NO and even more restricted for the orbital pair specific pair NO  density matrices.

\subsection{Strong pair classification}\label{sec:strong_pair}
The convergence of the LNO-CCSD(T) energies with the tightening of the pair energy threshold ($\varepsilon_\mathrm{w}$) is also investigated.
The restriction of the strong pair LMO list, controlled by this threshold, only starts to take effect for
 somewhat larger systems, preventing one from utilizing canonical DF-CCSD(T) calculations as reference. 
To overcome this difficulty, first, we recall that this approximation is the main source of error 
in the LMP2 correlation energy, which we benchmarked against DF-MP2 for a number of larger open-shell molecules containing 42--81 atoms.
Namely, the singlet-triplet gap of artemisinin, the vertical ionization energy of testosterone, and the hydrogen addition to vitamin E succinate 
 were investigated using the aug-cc-pVTZ basis set.\cite{ROLocalMP2} 
There, we found that the  energy differences converged
already at the default $\varepsilon_\mathrm{w}=10^{-5}$ E\textsubscript{h} LMP2 setting with
at most 0.03\% relative correlation energy and 0.05 kcal/mol energy difference deviations.

\begin{figure}
\centering
\includegraphics[width=\textwidth]{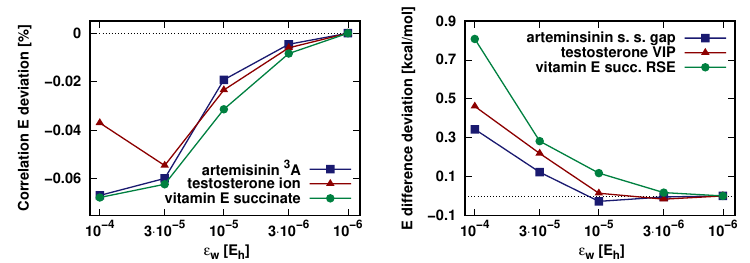}
\caption{Relative LNO-CCSD(T) correlation energy (left) and absolute energy difference (right) deviations.
A converged LNO-CCSD(T) calculation with $\varepsilon_\mathrm{w}=10^{-6}$ was used as reference.
See Section \ref{sec:strong_pair} for a description of the utilized systems and basis sets.}
\label{fig:wpairtol}
\end{figure}

To also assess the accuracy of LNO-CCSD(T) against an affordable reference on the same systems, we set all thresholds to their default value
except for the $\varepsilon_\mathrm{w}$ parameter, which we scan (see Figure \ref{fig:wpairtol}).
Therefore, as reference, the converged calculation with the very tight $\varepsilon_{\mathrm{w}}=10^{-6}$ E\textsubscript{h} can be taken.
With that, we find the relative correlation energy and absolute energy difference deviations
to converge quickly to zero also for LNO-CCSD(T)  as the pair energy threshold is tightened.
Note that the absolute energy difference considered for vitamin E succinate is about four (two) times larger than that for artemisinin (testosterone), meaning that the slightly larger energy difference deviations observed for vitamin E succinate 
 correspond to similarly high quality relative deviations for all systems.
The value of $\varepsilon_\mathrm{w}=10^{-5}$ E\textsubscript{h}, yielding highly satisfactory LNO-CCSD(T) results of
 less than 0.04\% correlation energy deviations and highest energy difference deviations of just above 0.1~kcal/mol, can be again safely chosen as default.
 Our strong pair threshold can be compared to the settings of other local approaches, and for example, matches 
the value employed in the TightPNO settings of the DLPNO methods.\cite{OS_DLPNO-CCSD(T)}
This value matches our closed-shell setting as well, where the restriction of the strong pair LMO list was successfully benchmarked 
for even larger molecules of 92--260 atoms against DF-MP2 references.\cite{LocalMP2}

\subsection{Approximate long-range spin polarization}
The accuracy of the distant spin polarization approximation 
introduced in Section \ref{sec:closed_shell_comparison} is evaluated on both local correlation energies and energy differences. 
Here, the reference energies are obtained from default LNO-CCSD(T) calculations without activating the distant spin polarization approximation.
Since the approximation is only active in the EDs that do not contain SOMOs, reasonably large systems should be chosen to activate and test this approach.
Results for three medium to large test systems of 81--179 atoms 
 are collected in Table \ref{tab:csapprox}. 
As shown in the last column of  Table \ref{tab:csapprox}, already at this relatively moderate system size,
 one half to two thirds of the EDs without SOMOs can be treated with the more efficient closed-shell formalism.
At the same time, all relative correlation energy errors are well below 10\textsuperscript{-4} \%, indicating only a marginal loss of accuracy.
The absolute and relative errors in the computed energy differences of about 0.1--0.4~cal/mol or up to  5.4$\cdot$10\textsuperscript{-4} \% are similarly negligible. It is also worthwhile noting that here, the RSE and bond breaking processes in the first and third 
lines involve also large closed-shell species, where this approximation does not have an analogue eliminating the possibility of error cancellation.
In conclusion, the errors introduced by the distant spin polarization approximation are substantially smaller than those brought in by the other local approximations employed.
These results are in line with the performance of the analogous approximation in our open-shell LMP2 approach, where we 
have also performed such tests up to 500-600 atoms. \cite{ROLocalMP2} Notably, in that size range, 80--90\% of the EDs are free 
of SOMOs and can be treated with the more efficient restricted algorithms.
\begin{table}
\caption{Accuracy of the long-range spin polarization approximation  compared to reference LNO-CCSD(T) correlation energies and energy differences obtained without this approximation.} 
\label{tab:csapprox}
\small
\setlength{\tabcolsep}{4pt}
\begin{tabular}{m{2.2cm}m{0.8cm}|cccccc}
&&\multirow{2}{*}{atoms}&\multirow{2}{*}{LMOs}&\multirow{2}{2.5cm}{\centering $E^{\mathrm{LNO-CCSD(T)}}$  \\ error [\%]}& \multicolumn{2}{c}{ error in energy difference}  &\multirow{2}{2.5cm}{\centering EDs without SOMOs [\%]} \\ \cline{6-7}
&&&&&[cal/mol]&[\%]\\
\hline 
\multicolumn{2}{l|}{vitamin E succinate} & 81 & 89 & $1.2\cdot 10^{-5}$& 0.44 & $1.1\cdot 10^{-4}$ & 54 \\
\multirow{2}{*}{FeC\textsubscript{72}N\textsubscript{2}H\textsubscript{100} \;\;}& $^5$A  & \multirow{2}{*}{175} & 205 & $3.3\cdot 10^{-5}$ & \multirow{2}{*}{0.22} &\multirow{2}{*}{$5.4\cdot 10^{-4}$} &54 \\
& $^3$A & & 204 &$3.5\cdot 10^{-5}$& & &54 \\
\multicolumn{2}{l|}{Cbl radical}            & 179 & 250 &$2.0\cdot 10^{-6}$ & 0.087 & $1.6\cdot 10^{-4}$ &68\\
\end{tabular}
\end{table} 


\subsection{Natural auxiliary functions} \label{sec:nafthr}
In this section, the effect of the NAF truncation is investigated on three systems of up to 81 atoms with various basis sets. These include
the vertical ionization energy of testosterone (cc-pVTZ),
 the hydrogen addition process of vitamin E succinate (aug-cc-pVTZ),
 and the singlet-triplet gap of diphenylcarbene (cc-pVQZ). 
Since the NAF approximation was found very effective in the analogous closed-shell context,\cite{LocalCC3,LocalCC4} and
the only difference compared to the latter is that here, both spin up and spin down integrals have to be expanded in the same NAF basis,
we expect very similar accuracy.

\begin{figure}
\centering
\includegraphics[width=\textwidth]{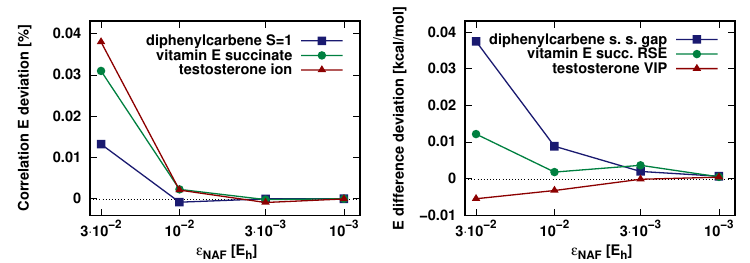}
\caption{Errors of LNO-CCSD(T) energies and energy differences computed with various NAF truncation thresholds. The reference energies are obtained from LNO-CCSD(T) calculations with $\varepsilon_\text{NAF} = 0$. Left panel: relative correlation energy differences, right panel: absolute deviations in energy differences. 
}
\label{fig:naf}
\end{figure}

To evaluate the performance of NAFs, LNO-CCSD(T) calculations with  default settings except for the scanned $\varepsilon_\mathrm{NAF}$ parameter
are compared against the reference obtained with $\varepsilon_\mathrm{NAF}=0$.
Figure \ref{fig:naf} shows the convergence of the correlation energy (left) and energy difference (right) deviations.
The correlation energy deviations converge almost completely already 
with $\varepsilon_{\mathrm{NAF}}=10^{-2}$ E$_\mathrm{h}$, while energy difference deviations approach the reference also almost completely monotonically.
Since the largest energy difference deviation is already below 0.01 kcal/mol with $\varepsilon_\mathrm{NAF}=10^{-2}$ E$_\mathrm{h}$, this value is chosen as the default,
again matching the default setting of  closed-shell LNO-CCSD(T). 


\section{Benchmarks for small and medium sized systems} \label{sec:statistics}
In this section, the cumulative effect of all local truncation thresholds employed at their default values is benchmarked against canonical DF-CCSD(T) references.
Statistical measures for the correlation energy and energy difference deviations are presented for three test sets 
containing radical stabilization energies, ionization potentials, and spin-state energies. 
The reference data utilized in this section is available in the Supporting Information.  

\subsection{Accuracy of local correlation energies}
Here, the accuracy of the open-shell LNO-CCSD(T) correlation energies is evaluated against reference DF-CCSD(T) values.
The investigated systems include 30 radicals (from the RSE30 test set), 21 cations from IP21, and the 12 aryl carbenes in their triplet state of the AC12 test set.

The statistical measures of the correlation energy deviations of the three test sets are collected in the third columns of Tables \ref{tab:rse30}--\ref{tab:ac12}.
For all test sets, the MAEs of the relative correlation energy deviations with the triple-$\zeta$ quality basis sets are below 0.05\%, with all errors not exceeding 0.11\%.
The highest triple-$\zeta$ deviation of 0.11\% is observed for the oxalic acid cation, where a large (0.35) single excitation amplitude of the reference DF-CCSD(T) solution suggests some multireference character.
Apart from this challenging system lying at the boundary of single reference CC's applicability, all triple-$\zeta$ local correlation energy deviations are below 0.1\%, 
which is highly satisfactory already with the default LNO-CCSD(T) settings. 
When moving to the quadruple-$\zeta$ and triple-$\zeta$--quadruple-$\zeta$ extrapolated [CBS(T,Q)] results of the RSE30 test set, 
one observes a slight increase in the MAE and MAX local correlation energy deviations.
It is worth noting that the largest relative error is observed in both cases for the methyl radical.
Due to its small size, the reference correlation energy values for this system are tiny, 
greatly increasing the relative correlation energy deviations even for the smallest of absolute errors.
Compared to the high quality of MAEs for the correlation energies, it is important to point out that their
STD is always as good as or, in a number of cases, 2--3 times smaller than the MAE 
for all three test sets. This indicates that one can expect excellent cancellation between the correlation 
energy errors upon computing various energy differences, as found in the next section. 
\begin{table}
\caption{Relative correlation energy deviations and absolute errors of the LNO-CCSD(T) reaction energies for 
radical stabilization energies of the RSE30 test set using the default thresholds.}
\begin{tabular}{cccc}
basis & error measure & {error in $E^{\mathrm{LNO-CCSD(T)}}$ [\%]}& error in RSE [kcal/mol] \\
\hline
\multirow{3}{*}{aug-cc-pV(T+d)Z}
 & MAX & 0.081 & 0.111 \\
 & MAE & 0.029 & 0.042 \\
 & STD & 0.034 & 0.041 \\  
\hline
\multirow{3}{*}{aug-cc-pV(Q+d)Z}
 & MAX & 0.195 & 0.136 \\
 & MAE & 0.099 & 0.037 \\
 & STD & 0.050 & 0.050 \\
\hline
\multirow{3}{*}{CBS(T,Q)} 
 & MAX & 0.292 & 0.173 \\
 & MAE & 0.173 & 0.050 \\
 & STD & 0.063 & 0.069 \\
\end{tabular}
\label{tab:rse30}
\end{table} 

\begin{table}
\caption{Relative correlation energy deviations and absolute errors of the LNO-CCSD(T) vertical ionization energies of the IP21 test set using the default thresholds.}
\begin{tabular}{ccccc}
basis & error measure &  error in $E^{\mathrm{LNO-CCSD(T)}}$ &  \multicolumn{2}{c}{error in IP}   \\
 & & [\%]   &  [eV] & [kcal/mol]  \\
\hline
\multirow{3}{*}{aug-cc-pV(T+d)Z}
 & MAX & 0.109 & 0.027  & 0.62  \\
 & MAE & 0.047 & 0.007  & 0.16  \\
 & STD & 0.058 & 0.009  & 0.21  \\  
\end{tabular}
\label{tab:ip21}
\end{table} 

\begin{table}
\caption{Relative correlation energy deviations and absolute errors of the LNO-CCSD(T) singlet-triplet energy gaps of the AC12 test set using the default thresholds.}
\begin{tabular}{cccc}
basis & error measure & {error in $E^{\mathrm{LNO-CCSD(T)}}$ [\%]}& error in S--T gap [kcal/mol] \\
\hline
\multirow{3}{*}{cc-pVDZ}
 & MAX & 0.073 & 0.337 \\
 & MAE & 0.041 & 0.106 \\
 & STD & 0.017 & 0.118 \\  
\hline
\multirow{3}{*}{cc-pVTZ}
 & MAX & 0.084 & 0.452 \\
 & MAE & 0.043 & 0.244 \\
 & STD & 0.021 & 0.159 \\
\end{tabular}
\label{tab:ac12}
\end{table} 

\subsection{Radical stabilization energies}\label{sec:rse30}
The RSE30 test set contains radical stabilization reactions of the form:
\begin{equation}
\mathrm{R} \hskip -0.1cm - \hskip -0.1cm \mathrm{H} + \mathrm{CH}_3 \, \cdot \longrightarrow \mathrm{R} \cdot + \, \mathrm{CH}_4 \ ,
\label{eqn:radical_stabilization}
\end{equation}
where $\mathrm{R} \, \cdot$, stands for the different radicals containing C, N, O, F, P, and S atoms \cite{PNO-UCCSD}.
The deviations of LNO-CCSD(T) radical stabilization energies from reference DF-CCSD(T) ones are collected in the last column of Table \ref{tab:rse30}.
With MAEs of at most 0.05 kcal/mol and MAX errors below 0.2 kcal/mol, even for the CBS extrapolated energy differences, the results are highly satisfactory.
Here, the 0.050 \% and 0.063\% STDs of the quadruple-$\zeta$ and CBS(T,Q) correlation energies, respectively,
are important indicators to explain that the triple-$\zeta$, quadruple-$\zeta$,  and CBS(T,Q) RSE values are of the same high quality. 
One might also compare the results obtained with the aug-cc-pV(T+d)Z basis to those of the PNO-UCCSD(T)-F12 \cite{PNO-UCCSD(T)} method, calculated for the same RSE30 test set.
Naturally, the PNO-UCCSD(T)-F12  and the LNO-CCSD(T) are not directly comparable as the former
 contains additional explicitly correlated terms.
Nonetheless, the LNO-CCSD(T) results appear to be comparable to or slightly better than 
the PNO-UCCSD(T)-F12 results with defaults settings providing 0.076 kcal/mol RMS, 0.064 kcal/mol MAE, and 0.192 kcal/mol MAX errors with the 
aug-cc-pVTZ basis set.\cite{PNO-UCCSD(T)} It is always important to keep in mind at such comparisons that both 
the PNO-UCCSD(T)- and the LNO-CC-type methods can be converged to the same local approximation free limit, and here, we 
compare only the results obtained with the default settings of each method determined according to considerations that are not identical for the two methods.


\subsection{Vertical ionization potentials}
The statistical measures of the vertical ionization potentials of the IP21 test set are shown in Table \ref{tab:ip21}.
Considering the fact that the IPs represent quite large energy differences in the range of 8--14 eV (184--323 kcal/mol), the mean average error of 0.007 eV (0.16~kcal/mol) is excellent. 
The MAX IP deviation of 0.027 eV can be attributed to the benzoquinone cation, the reference DF-CCSD(T) solution of which exhibits a moderately large maximum singles amplitude of 0.15.
Removing this outlier, the MAX IP error drops to below 0.02 eV (0.04 kcal/mol), and the standard deviation also decreases to 0.007 eV (0.16~kcal/mol).
The PNO-UCCSD(T)-F12 method achieves comparable or slightly more accurate results for this test set with its default 
settings and the aug-cc-pVTZ basis set
(0.17 kcal/mol RMS, 0.11 kcal/mol MAE, and 0.50~kcal/mol MAX).\cite{PNO-UCCSD(T)}


\subsection{Singlet--triplet energy gaps}
Finally, the singlet--triplet spin state energy gaps of the aryl carbenes are benchmarked, and the results are collected in the last column of Table \ref{tab:ac12}.
The MAE of 0.24 kcal/mol for the cc-pVTZ basis is somewhat larger than those for the other two test sets but is still only a fraction of a kcal/mol
marking chemical accuracy.
The slightly increased errors are at least partly explained by the fact that the system sizes of this test set are larger than that of the other two, with the average AC12 system containing more than twice as many heavy atoms as either RSE30 or IP21.
To put the results in perspective, let us consider that the basis set incompleteness error of canonical CCSD(T) for the AC12 test set 
found to be 2.74 and 0.98 kcal/mol, respectively, for the cc-pVDZ and cc-pVTZ bases,\cite{aryl_carbenes} which are 
significantly larger than the MAX local errors reported here.



\section{Performance and computational requirements for larger systems}\label{sec:large_calcs}
The current capabilities of the open-shell LNO-CCSD(T) method are illustrated on large-scale calculations on four three-dimensional systems
 containing 175--601 atoms (see Table~\ref{large_calcs}).
Of these systems, the Cbl radical and the FeC\textsubscript{72}N\textsubscript{2}H\textsubscript{100} complex represent 
the higher end of the typical size range, e.g., in homogeneous catalysis applications (cca.~175 atoms), while 
the bicarbonate or DAAO models of 565--601 atoms are in the typical size range of 
quantum systems in biochemical applications, 
for example,  when modeling  proteins with active centers in a quantum mechanics/molecular mechanics (QM/MM) framework.
 The Cbl radical and the FeC\textsubscript{72}N\textsubscript{2}H\textsubscript{100} complex 
 pose a formidable challenge for local correlation methods
due to the spin polarized transition metal atoms near their centers.
This leads not only to  high strong pair ratios of 22--26\% but also to LISs that are somewhat more extended 
than those of the two protein models. 
It is therefore not completely surprising to see that the wall clock run times of the domain CCSD(T) calculations for the 
transition metal complexes somewhat exceed even those for PSII bicarbonate or the singlet DAAO species. 
Another reason for the unusually long CCSD(T) run time for the FeC\textsubscript{72}N\textsubscript{2}H\textsubscript{100} system is its moderate multireference character \cite{SparseMapsNEVPT2}, which hinders the convergence of the CCSD iterations, resulting in up to 27 iterations performed in particular domains, compared to the usual 9--11 iterations.

\begin{table}[h!]
\caption{Average (maximum) domain sizes, orbital space dimensions, DF-HF and correlation energies (in E\textsubscript{h}), wall-clock times
(in hours)\textsuperscript{\emph{a}}, and memory requirements (in GB) for LNO-CCSD(T) computations of large molecules.}
\footnotesize
\label{large_calcs}
\begin{tabular}{lccccc}
molecule&FeC\textsubscript{72}N\textsubscript{2}H\textsubscript{100} & Cbl radical & bicarbonate & \multicolumn{2}{c}{DAAO} \\
\hline 
atoms               & 175          & 179          & 565         & \multicolumn{2}{c}{601} \\
LMOs                & 205          & 250          & 788         & 837 & 838 \\
SOMOs               & 4            & 1            & 2           & 0 & 2 \\
AO basis            & def2-TZVP    & def2-TZVP    & def2-TZVP   & \multicolumn{2}{c}{def2-TZVP} \\
basis functions     & 2939         & 3369         & $10 \, 560$ & \multicolumn{2}{c}{$11 \, 006$}\\
auxiliary functions & 7306         & 8379         & $26 \, 064$ & \multicolumn{2}{c}{$27 \, 071$}\\
strong pairs [\%]   & 26           & 22           & 6.8 & 5.9   & 5.9 \\
atoms in ED         & 116 (171)    & 116 (169)    & 132 (294)   & 124 (268)   & 137 (353) \\
AOs in ED           & 2129 (2915)  & 2346 (3278)  & 2634 (5953) & 2453 (5408) & 2700 (6943)\\ 
LMOs in ED          & 54 (111)     & 55 (115)     & 55 (112)    & 50 (113)    & 51 (184)   \\
PAOs in ED          & 1045 (1949)  & 1023 (1829)  & 975 (2037)  &  866 (1884) &  905 (2932) \\
Occupied LNOs       & 32 (64)      & 31 (74)      & 29 (58)     &  26 (65)    &  27 (100)   \\
Virtual LNOs        & 131 (268)    & 125 (272)    & 121 (250)   &  112 (269)  &  113 (449)  \\
\hline
type of reference   & ROHF         & ROHF         & QRO (UHF)   & RHF         & ROHF\\ 
DF-HF energy        & -4156.159945 & -5878.796625 & -15197.85043\textsuperscript{\emph{c}} & -14740.93979 & -14740.90403\\
LNO-CCSD(T) energy  & -13.3715     & -17.6478     & -56.0763    & -59.1499    & -59.1421 \\
\hline
HF (1 iteration)    & 0.50          &  0.72        & 3.1\textsuperscript{\emph{b}} & 2.5\textsuperscript{\emph{b}}  & 2.6\textsuperscript{\emph{b}} \\   
orbital localization        & $<0.01$ & $<0.01$ & 0.12 & 0.06 & 0.05 \\
pair energies       & 0.03         & 0.04         & 0.17        & 0.09        & 0.19 \\
LMP2 in EDs         & 0.74         & 0.97         & 1.1         & 0.67        & 1.93 \\     
integral trf. to LNOs       & 3.0          & 4.0          & 11          & 10          & 13  \\  
CCSD(T) in LISs & 75           & 55           & 29          & 10          & 76 \\  
total LNO-CCSD(T)   & 85           & 67           & 57          & 33          & 106 \\  
\hline
memory req.         & 12           & 18           & 17          & 7.1         & 88 \\ 
\end{tabular}

\raggedright{\textsuperscript{\emph{a}}  Using two 20-core, 2.4 GHz Intel Xeon Gold 6148 CPUs} \\
\raggedright{\textsuperscript{\emph{b}}  Using the default local fitting domain size. The final iteration with larger fitting domains took
about 3.5--4.8 times longer.} \\
\raggedright{\textsuperscript{\emph{c}}  DF-HF energies calculated with semi-canonical QRO orbitals.} \\
\end{table}

Regarding the other contributions to the run time, it is reassuring that the formally
 cubic scaling orbital localization and the quadratically scaling pair energy evaluation take negligible time, even for the largest systems.
However, the cubic-scaling local DF-HF computations can take time comparable to the LNO-CCSD(T) correlation energy computation
for the large protein models, in accord with our experience with even larger proteins of 1000--2000 atoms 
and our restricted LNO-CCSD(T) implementation.\cite{LocalCC3,LocalCC4}
Moreover, as observed on the singlet and triplet DAAO models, the time required for the 
integral transformation to the LNO basis increases by only up to 30\% compared to the closed-shell algorithm,
which increase is mainly attributed to the larger domain sizes within the triplet system.
As discussed in Section \ref{sec:closed_shell_comparison}, an unrestricted  implementation of the 
open-shell integral transformation would have twice the computational cost of the closed-shell analogue. 
However, due to the restricted intermediate basis 
and the distant spin polarization approximation,
 the computational costs of the most expensive integral transformation steps approach those of the closed-shell algorithm.
This small overhead of the open-shell integral transformations means that, while this step is one of the rate determining parts of the closed-shell algorithm (30\% of the singlet DAAO run time), its significance is much reduced in the open-shell case (12\% of the triplet DAAO run time).

Looking at the average (maximum) ED sizes of 116--137 (169--353) atoms, at least for the large protein models, 
the individual domain sizes are saturated, and we are getting close to the linear scaling regime of the ED and LIS computation parts of the LNO-CCSD(T) method. It is also clear that performing hundreds of MP2 energy computations 
for these domains is well beyond the capabilities of any conventional MP2 implementation. 
These ED MP2 computations are only feasible with a thoroughly optimized,
Cholesky-decomposition- or
Laplace-transform-based algorithm and extensive utilization of local approximations.\cite{LocalMP2,ROLocalMP2}
In these (necessarily) large EDs, we can exploit the locality of the LMOs, use local AO and DF domains, 
and redundancy free MP1 amplitude expressions.
As a result of  these enhancements, the MP2 energy computation in the largest ED
with over 180 LMOs and almost 3000 PAOs takes less than an hour.
Turning to the average LIS sizes, the necessity and efficiency of the LNO compression is evident.
The LISs contain somewhat over half as many occupied and around 6--8 times fewer virtual orbitals after the truncation of the ED basis dimensions.
Without this compression of the correlated subspaces, the evaluation of hundreds of domain CCSD(T) energies would clearly be unfeasible.

\begin{figure}
\centering
\includegraphics[width=\textwidth]{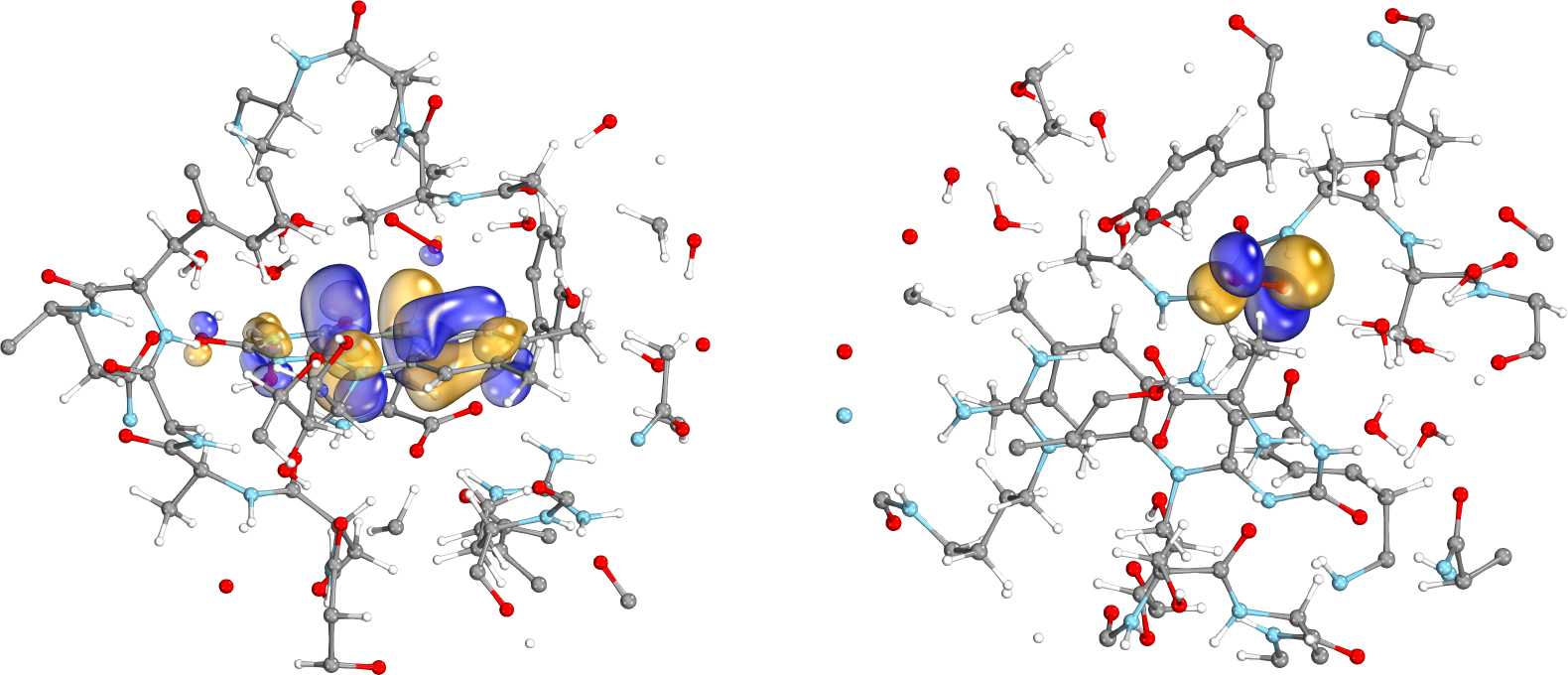}
\caption{\textsc{d}-amino-acid oxidase model with its two singly occupied molecular orbitals depicted with blue and yellow surfaces.
The left panel shows the highly delocalized orbital spreading over the entire flavin adenine dinucleotide, while the other singly occupied orbital localized almost exclusively on the peroxide radical can be seen on the right.}
\label{fig:daao_large_orbital}
\end{figure}

It is interesting to inspect the DAAO systems more closely due to the availability of very similar closed-shell and open-shell computations.
 Here, one immediately notices the surprisingly large maximum domain size of the triplet DAAO model in the last column of Table~\ref{large_calcs}.
Containing over 350 atoms, 100 occupied, and 449 virtual LNOs, the size of this domain is unprecedented
 in LNO-CCSD(T) calculations performed with the default threshold values.
The central LMO of this domain is one of the SO LMOs of the system, depicted on the left panel of Figure \ref{fig:daao_large_orbital}.
This orbital spreads across the entire FAD moiety and, to a small extent, even spills onto the peroxide molecule.
The large spatial extent of this orbital means that it forms an unusually large number of 
 strong pairs with other LMOs. 
This in turn leads to a large number of ED atoms and finally, to large LIS orbital dimensions. 

Beyond the inherently delocalized electronic structure of the reduced FAD,
 another reason for the extreme spatial extent of this SO LMO is to be found in the restricted nature of the employed orbital localization scheme.
During localization, the SOMOs are mixed only among themselves, in order to avoid the splitting of the spin up and spin down occupied orbitals.
The triplet state of the DAAO model contains merely two SOMOs, but at the same time the singly occupied subspace  spreads across
 the entire FAD moiety and the peroxide molecule.
Due to the large spatial extent of the SO subspace and the small number of orbitals spanning it, 
it is not possible to localize the SO LMOs better in a restricted scheme.
A (partial) remedy could be to employ a spin unrestricted localization scheme,
 where the two SOMOs would be allowed to mix with all other spin up orbitals.
This  could probably lead to orbitals more localized than the highly delocalized SOMO in question
 at the cost of introducing spin splitting to the LMOs, doubling the number of domains in all computations.  
Despite the complex electronic structure and extreme domain size, the LNO-CCSD(T) energy contribution of this delocalized
 central LMO can still be evaluated on a single CPU. 
This would not be possible without efficient local approximations, such as the compact NAF and LNO orbital spaces, 
the highly optimized CCSD(T) code featuring integral direct 
 evaluation of four-center more-than-two-external ERIs, and the redundancy free Laplace-transform (T) formulation of LNO-CCSD(T). 
The broader significance of the present triplet DAAO LNO-CCSD(T) computation is that it represents (i) the largest   CCSD(T) computation 
in the literature with over 600 atoms and more than twice as many orbitals as in the  def2-SVP level  DLPNO-based PSII bicarbonate computation 
of Kumar et al.\cite{DLPNO_CCSD(T)-F12_OS}, (ii) probably the most complicated system studied so far concerning the issues with delocalized SOMOs, illustrating that such problematic situations can also be handled with highly-optimized implementations. 

It is also encouraging to see the comfortably manageable minimal memory requirements in the last line of Table~\ref{large_calcs}, 
despite the complexity of the investigated systems and the employed triple-$\zeta$ quality basis set.
Less than 20 GB of minimal memory  is required for most systems in  Table~\ref{large_calcs},
 while with about 90 GB, LNO-CCSD(T) was feasible even for the most challenging triplet DAAO system.
Such low memory requirements are the result of the highly optimized integral evaluation and transformation routines and the tiled computation of the most memory intensive CCSD(T) terms, which removes the need to store any four-center ERI arrays
 with more than two external indices in the memory.
Additionally to the small minimal memory requirements, the wall-clock run times of 1.5--4.5 days 
also indicate that LNO-CCSD(T) calculations for open-shell systems
containing several hundred atoms,
 at least with reasonable triple-$\zeta$ quality basis sets and the default settings, 
 have become routine tasks even on a  single  many-core CPU with easily accessible amount of memory.

Lastly, the reaction energies and energy differences obtained with the LNO-CCSD(T) method for the above
 discussed four large examples are collected in Table \ref{tab:large_e_diff}.
Comparing the estimates of the energy differences obtained at the LMP2 and LNO-CCSD(T) levels of theory, one observes significant differences for all applications.
The deviation of the relative energy estimates between these two methods is smallest for the DAAO oxidation and dAdoCbl formation processes, however, with 2.1--2.4 kcal/mol, this deviation is still clearly outside of
the range of chemical accuracy typically defined to be 1 kcal/mol.
Moreover, for the quintet--triplet energy differences of the two iron complexes, the disagreement between the two methods increases to 6.6--6.8 kcal/mol.
These results are in accord with the observations emerging more and more often in the literature, enabled by large-scale CCSD(T) benchmarks, namely that second order perturbation theory can often be insufficient to accurately describe such challenging extended systems and processes. This also verifies the need for higher order and more reliable methods.  
\begin{table}
\caption{Contributions to reaction energies and spin-state gaps in kcal/mol for the four largest examples evaluated with the def2-TZVP basis.
$\Delta E^{\mathrm{HF}}$ -- DF-HF,
$\Delta E^{\mathrm{LMP2}}$ -- LMP2 correlation contribution,
$\Delta E^{\mathrm{LNO-CCSD(T)}}$ -- LNO-CCSD(T) correlation contribution,
$\Delta E_{\mathrm{total}}^{\mathrm{LNO-CCSD(T)}}$ -- total LNO-CCSD(T).}
\label{tab:large_e_diff}
\begin{tabular}{l|cccc}
  & $\Delta E^{\mathrm{HF}}$ & $\Delta E^{\mathrm{LMP2}}$ & $\Delta E^{\mathrm{LNO-CCSD(T)}}$ & $\Delta E_{\mathrm{total}}^{\mathrm{LNO-CCSD(T)}}$ \\
\hline
FeC\textsubscript{72}N\textsubscript{2}H\textsubscript{100} $^5$A -- $^3$A & 57.56 & -10.57 & -17.39 & 40.17 \\
bicarbonate $^5$A -- $^3$A & 52.67 & -12.07 & -18.70 & 33.97 \\
Cbl + Ado $\to $ dAdoCbl & 50.41 & -102.47 & -104.82 & -54.41 \\
DAAO oxidation & 22.44 & 7.01 & 4.88 & 27.32 \\
\end{tabular}
\end{table} 


\section{Summary and conclusions}

A high-spin open-shell extension of the asymptotically linear-scaling local natural orbital (LNO) based CCSD(T) method is presented.
The efficiency of the open-shell algorithm approaches that of our closed-shell LNO-CCSD(T) method \cite{LocalCC3} for large molecules due to the utilization of restricted open-shell Hartree--Fock (ROHF) or Kohn--Sham (ROKS) reference determinants, the use of restricted open-shell intermediate basis sets, and a novel approximation of long-range spin-polarization effects at the CCSD(T) level. 
For compact molecules, where the domain CCSD(T) computations are rate-determining, the open-shell LNO-CCSD(T) method requires at least 
2--3 times more operations and data than the closed-shell variant, just as for alternative local CCSD(T) approaches.
The presented method is identical to its closed-shell counterpart when closed-shell systems are considered, enabling the combined use of these algorithms to efficiently compute consistent energy differences between open- and closed-shell species.

The proposed approach efficiently extends and generalizes the optimized algorithms developed for our other local correlation methods to the open-shell LNO-CCSD(T) case \cite{LocalMP2,LaplaceT,LocalCC3,LocalCC4,ROLocalMP2}, for example, the algorithms for domain constructions, auxiliary basis compression, integral transformation, and LNO construction.
Its implementation is integral-direct, memory- and disk use economic, 
and OpenMP-parallel, although the scaling of the current version above 10--20 cores should be improved.
The local approximations are defined without relying on empirical parameters, such as real-space cutoffs of fragment definition, and adapt  to the complexity of the wave function by construction.
Additionally, a highly accurate open-shell local MP2 correlation energy is also obtained for free as part of the LNO-CCSD(T) calculation.
The open-shell LNO-CCSD(T) code will be made available open-access for academic use in a forthcoming  release of the 
\textsc{Mrcc} program suite.\cite{MRCC,mrcceng3}

The effect of the local approximations and the corresponding thresholds are profiled on both energy differences and correlation energies.
Systematically tightening these thresholds, satisfactory convergence towards the exact CCSD(T) energies is observed.
The errors caused by the local approximations using the default threshold values are benchmarked on diverse test sets of radical stabilization energies, vertical ionization potentials, and singlet--triplet energy gaps. 
For systems where the exact CCSD(T) reference is accessible,
mean absolute errors are found to be below 0.25 kcal/mol, while maximum errors do not exceed 0.7 kcal/mol and are below 0.5 kcal/mol for all but the most challenging systems.

The performance of the open-shell LNO-CCSD(T) algorithm is demonstrated on systems of 175--601 atoms using triple-$\zeta$ quality basis sets.
Among these, a triplet state of the O\textsubscript{2} reduction process via a \textsc{d}-amino acid oxidase model of 601 atoms is especially challenging, not only due to its unprecedented size with over 11,000 basis functions, but also because of a singly-occupied molecular orbital delocalized over about 20 atoms.
The complicated triplet electronic structure of the 565-atom model of the bicarbonate protein in photosystem II is also investigated using about twice as many basis functions as in a previous local CCSD study.\cite{DLPNO_CCSD_OS}
Each of these calculations is feasible using less than 100 GB of memory, on a single node of 20--40 cores within cca.~2.5--4.5 days of wall clock run time.
These results demonstrate that the presented LNO-CCSD(T) algorithm extends the applicability range of open-shell CCSD(T) calculations up to biochemical systems of 500-600 atoms using reasonable triple-$\zeta$ basis sets and commodity hardware.
Having the ROHF/ROKS/QRO reference at hand, which can become relatively costly in this size range 
the open-shell LNO-CCSD(T) method should scale to even larger systems/basis sets, which approaches the capabilities of closed-shell LNO-CCSD(T), currently standing at about 1000-2000 atoms and 45,000 basis functions \cite{LocalCC3,LocalCC4}.


\appendix
\section*{Appendix: Derivation of virtual density matrix fragments}\label{sec:appendix_dm}


In this Appendix, the orbital notation  of the  main text is simplified using 
$i$,$j$,$k$\ldots and $a$,$b$,$c$\ldots indices for the occupied and  virtual (semi-)canonical orbitals, respectively.
Again, lowercase letters stand for spin up, uppercase letters label spin down orbitals, while MP1 amplitudes are denoted by $t$.
Moreover, expressions only for the spin up density matrix fragments are shown since their spin down counterparts can be
 obtained by mirroring all spins of the expressions.

The spin up  virtual second-order density matrix reads as
\begin{equation}
D_{ab}=\frac{1}{2}\left(\sum_{cij}{t^{ac}_{i j} t^{bc}_{i j}}+\sum_{CiJ}{t^{aC}_{i J} t^{bC}_{i J}}+\sum_{CIj}{t^{aC}_{I j}t^{bC}_{I j}}\right) \, .
\label{eqn:canonical_mp2_dm}
\end{equation}
After transforming one of the occupied orbitals, say the $i$ and $I$ indices, to the LMO basis, 
one has three different choices to continue the derivation. 
If we first restrict the summations to the central LMO index, we arrive at the virtual density matrix 
 expression used in the presented open-shell LNO-CCSD(T) algorithm (eq \ref{eqn:virt_can_density}).
The other two forms can be obtained by recognizing that the last two terms of eq \ref{eqn:canonical_mp2_dm} can be combined together
in two ways.
One can either switch the order of the occupied indices of the last term of eq \ref{eqn:canonical_mp2_dm}
and rename them to $I \to J$, $j \to i$, yielding:
\begin{equation}
\begin{split}
D_{ab}=
&\frac{1}{2}\left(\sum_{cij}{t^{ac}_{i j} t^{bc}_{i j}}+\sum_{CiJ}{t^{aC}_{i J} t^{bC}_{i J}}+\sum_{CIj}{t^{aC}_{I j}t^{bC}_{I j}}\right)= \\
&\frac{1}{2}\left(\sum_{cij}{t^{ac}_{i j} t^{bc}_{i j}}+\sum_{CiJ}{t^{aC}_{i J} t^{bC}_{i J}}+\sum_{CiJ}{t^{aC}_{i J}t^{bC}_{i J}}\right) =
\frac{1}{2} \sum_{cij}t^{ac}_{ij}t^{bc}_{ij} + \sum_{CiJ}t^{aC}_{iJ}t^{bC}_{iJ}\, .
\end{split}
\label{eqn:first_form}
\end{equation}
Here, we exploited  that exchanging the occupied indices of the antisymmetric MP1 amplitudes introduces two sign changes that cancel each other ($t^{aC}_{Ij} = - t^{aC}_{jI}$).
Transforming the $i$ index to the LMO basis and restricting its summation to the central LMO, 
one  obtains a virtual density matrix expression different from the first one:
\begin{equation}
\begin{split}
D_{ab}^\mathcal{I}=
\frac{1}{2} \sum_{cj}t^{ac}_{i^\prime j}t^{bc}_{i^\prime j} + \sum_{CJ}t^{aC}_{i^\prime J}t^{bC}_{i^\prime J}\, .
\end{split}
\label{eqn:third_form}
\end{equation}
However, in this form the central LMO with spin down occupation does not contribute to the spin up virtual density matrix fragment at all.
For the last form, one can perform the exchanging and renaming of the occupied indices on the second term of eq \ref{eqn:canonical_mp2_dm}, yielding:
\begin{equation}
\begin{split}
D_{ab}=
&\frac{1}{2}\left(\sum_{cij}{t^{ac}_{i j} t^{bc}_{i j}}+\sum_{CiJ}{t^{aC}_{i J} t^{bC}_{i J}}+\sum_{CIj}{t^{aC}_{I j}t^{bC}_{I j}}\right)= \\
&\frac{1}{2}\left(\sum_{cij}{t^{ac}_{i j} t^{bc}_{i j}}+\sum_{CIj}{t^{aC}_{I j} t^{bC}_{I j}}+\sum_{CIj}{t^{aC}_{I j}t^{bC}_{I j}}\right) = 
\frac{1}{2}\sum_{cij}t^{ac}_{ij}t^{bc}_{ij} + \sum_{CIj}t^{aC}_{Ij}t^{bC}_{Ij}\, .
\end{split}
\label{eqn:second_form}
\end{equation}
If this form were to be used after restricting the summations to the central LMO index, both the central LMO with spin up and spin down occupation  would contribute to the spin up density matrix. However, the contribution of the spin down central LMO would be weighted twice as high as that of the spin up central LMO.


\begin{suppinfo}

See the Supporting Information for reference DF-CCSD(T) energies of the RSE30, IP21, and AC12 test sets.

\end{suppinfo}

\begin{acknowledgement}
The authors are grateful for the financial support from the ERC Starting Grant No. 101076972, ``aCCuracy'',
the  National Research, Development, and Innovation Office (NKFIH, Grant No. FK142489 and KKP126451),
 the J\'anos Bolyai Research Scholarship of the Hungarian Academy of Sciences,
\'UNKP-23-5-BME-408 New National Excellence Program of the Ministry for Culture and Innovation sourced from the NKFIH fund,
 and the computing time granted on the Hungarian HPC Infrastructure
at NIIF Institute, Hungary.
The research reported in this paper is part of project BME-EGA-02,
implemented with the support provided by the Ministry of Innovation and
Technology of Hungary from the National Research, Development and
Innovation Fund, financed under the TKP2021 funding scheme.
\end{acknowledgement}


\providecommand{\latin}[1]{#1}
\providecommand*\mcitethebibliography{\thebibliography}
\csname @ifundefined\endcsname{endmcitethebibliography}
  {\let\endmcitethebibliography\endthebibliography}{}

\begin{figure}[h!]
\begin{center}
\begin{tocentry}
\includegraphics{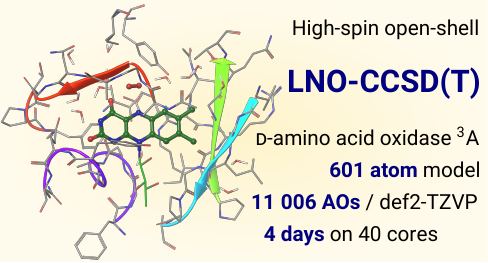}
\end{tocentry}
\end{center}
\end{figure}

\end{document}